\documentclass[twocolumn,times]{aastex63}
\usepackage{amsmath}
\usepackage{lipsum}
\usepackage{gensymb}
\usepackage{longtable}

\received{August 26, 2020}
\revised{October 8, 2020}
\accepted{October 10, 2020}
\submitjournal{AJ}

\shortauthors{Nguyen, De Rosa, and Kalas}

\graphicspath{{./}{figures/}}

\begin{document}

\title{First detection of orbital motion for HD 106906 b:  A wide-separation exoplanet on a Planet Nine-like orbit}

\correspondingauthor{Meiji M. Nguyen}
\email{meiji274@berkeley.edu}

\author[0000-0002-9350-4763]{Meiji M. Nguyen}
\affiliation{Department of Astronomy, University of California, Berkeley, CA 94720, USA}

\author[0000-0002-4918-0247]{Robert J. De Rosa}
\affiliation{European Southern Observatory, Alonso de C\'{o}rdova 3107, Vitacura, Santiago, Chile}

\author[0000-0002-6221-5360]{Paul Kalas}
\affiliation{Department of Astronomy, University of California, Berkeley, CA 94720, USA}
\affiliation{SETI Institute, Carl Sagan Center, 189 Bernardo Ave.,  Mountain View, CA 94043, USA}
\affiliation{Institute of Astrophysics, FORTH, GR-71110 Heraklion, Greece}

\begin{abstract}
HD 106906 is a 15 Myr old short-period (49\,days) spectroscopic binary that hosts a wide-separation (737 au) planetary-mass ($\sim11\,M_{\rm Jup}$) common proper motion companion, HD 106906 b. Additionally, a circumbinary debris disk is resolved at optical and near-infrared wavelengths that exhibits a significant asymmetry at wide separations that may be driven by gravitational perturbations from the planet. In this study we present the first detection of orbital motion of HD 106906 b using Hubble Space Telescope images spanning a 14 yr period. We achieve high astrometric precision by cross-registering the locations of background stars with the Gaia astrometric catalog, providing the subpixel location of HD 106906 that is either saturated or obscured by coronagraphic optical elements. We measure a statistically significant $31.8\pm7.0$ mas eastward motion of the planet between the two most constraining measurements taken in 2004 and 2017. This motion enables a measurement of the inclination between the orbit of the planet and the inner debris disk of either $36_{-14}^{+27}$\,deg or $44_{-14}^{+27}$\,deg, depending on the true orientation of the orbit of the planet. There is a strong negative correlation between periastron and mutual inclination; orbits with smaller periastra are more misaligned with the disk plane. With a periastron of $510_{-320}^{+480}$ au, HD 106906 b is likely detached from the planetary region within 100 au radius, showing that a Planet Nine--like architecture can be established very early in the evolution of a planetary system.
\end{abstract}

\section{Introduction}
\label{sec:intro}
Massive gas giants have been imaged in young stellar systems that still host large circumstellar disks, both at the gas-rich protoplanetary disk phase and the later debris disk phase. These planets undoubtedly play a significant role in the dynamical history of these systems, from carving out gaps in protoplanetary disks (e.g., \citealp{Keppler:2018dd}) to causing spiral arms, warps, and other perturbations (e.g., \citealp{Mouillet:1997ib,Oh:2016iu}). Perturbations within resolved images of protoplanetary and debris disks have also been used as indirect evidence of the presence of planets (e.g., \citealp{Partnership:2015cg,Dong:2015cp,Esposito:2016be,Ren:2020jq}), although alternative mechanisms do exist (e.g., \citealp{LorenAguilar:2015fa,Okuzumi:2016cd}). External perturbers have also been invoked in several examples, either by bound companions (e.g., \citealp{Rodriguez:2018hy,Wagner:2018dga}) or transient encounters with other stars within the birth cluster (e.g., \citealp{reche09a}). There are very few examples of systems where both the dynamical effect of an external perturber and its orbit can be determined, typically in cases where the perturber is a stellar companion. Finding additional examples will help further our understanding of their influence on the planet formation process.

\object{HD 106906} is a 15\,Myr old spectroscopic binary system \citep{Rodet:2017hr,DeRosa:2019ie} comprising two similar mass F-type main-sequence stars located in the Lower Centaurus Crux region of the Scorpius--Centaurus OB association 103 pc from the Sun \citep{Pecaut:2016fu}. An inner planetary system extending to roughly 50 au radius is inferred from a central dust clearing discovered within a near edge-on debris disk directly imaged at infrared wavelengths \citep{Kalas:2015en,Lagrange:2016bh}. Beyond the detected debris disk lies a directly imaged 11\,M$_{\rm Jup}$ planet, HD 106906 b, at a projected separation of 737\,au \citep{Bailey:2014et}. The dynamical evolution of the system is mysterious because the planet's projected position is $\sim$ 21\,deg away from the disk midplane, while the outer disk itself is highly asymmetric at optical wavelengths \citep{Kalas:2015en}. Specifically, the eastern side of the disk is vertically thin and extended to over 550\,au radius, while the western side of the disk is vertically thick and radially extended to 370\,au radius. Debris disks can be distorted by stellar flybys (e.g., \citealp{larwood01a}) and \citet{DeRosa:2019ie} discovered that HIP 59716 (F5V) and HIP 59721 (G9V) passed within $\sim$0.7 pc of HD 106906 2--3 Myr ago in a near coplanar encounter geometry. However, a closest approach distance $<$0.05 pc is needed to modify the orbits of either the planet or disk material \citep{rodet19a}. Theoretical work shows that the planet could gravitationally distort the disk if it has an eccentric orbit inclined to the disk midplane \citep{Jilkova:2015jo,Nesvold:2017ho, Rodet:2017hr}, inviting comparison to how Planet Nine may be aligning the eccentric orbits of detached Kuiper belt objects (KBOs) like Sedna \citep{trujillo14a,Batygin:2016efa}. However, obtaining an empirical measurement for the orbit of HD 106906 b is essential for validating the theoretical models.

In this paper we present an analysis of archival Hubble Space Telescope (HST) observations of the HD 106906 system where we measured significant orbital motion of the HD 106906 b planet over a 14 yr baseline. The observations and initial data analysis are described in Section~\ref{sec:obs} and our procedure to measure the Gaia sources within each image is described in Section~\ref{sec:astrometry}. The model we used to estimate the position of the HD 106906 star within each image is described in Section~\ref{sec:astro_model}, and its application to HD 106906 to precisely measure relative astrometry between the star and planet is described in Section~\ref{sec:application}. We present the first constraints on the orbit of the planet in Section~\ref{sec:results}, and discuss these results in Section~\ref{sec:discussion}.

\section{Observations and Data Reduction}
\label{sec:obs}
\begin{deluxetable*}{ccccccccc}
\tabletypesize{\normalsize}
\tablecaption{Observation Log\label{tbl:log}}
\tablehead{
\colhead{Date} &  \colhead{Instrument} & \colhead{Filter} & \colhead{Pivot Wavelength} & \colhead{Frames} & \colhead{Exposure Time} & \colhead{\texttt{ORIENTAT}\tablenotemark{$^a$}} &
\colhead{Plate Scale\tablenotemark{$^b$}} & \colhead{Proposal}\\
& & & (nm) & & (s) & (deg) & (mas\,px$^{-1}$)}
\startdata
2004 Dec 1 & ACS & F606W & 588.6 & 1 & 2$\times$0.1041 & 80.95 & 25.00 & GO-10330\\
            &     &       & & 1 & 2$\times$35     & & \\
            &     &       & & 1 & 2$\times$1250   & & \\ 
2016 Jan 29 & WFC3 & F127M & 1274 & 8 & 2$\times$66.4\tablenotemark{$^c$} & $-165.50$, $162.50$ & 128.25 & GO-14241\\
& & F139M & 1384 & 6 & 3$\times$88.4 & & &\\
& & F153M & 1532 & 6 & 3$\times$66.4 & & &\\
2017 Feb 24 & STIS & Clear & 5739 & 9 & 350.0\tablenotemark{$^d$} & $-84.84$, $-69.94$, $-54.94$  & 50.77 & GO-14670\\
2018 Jun 7 & WFC3 & F127M & 1274 & 28 & 2$\times$66.4\tablenotemark{$^c$} & $-58.50$, $-33.50$ & 128.25 & GO-14241\\
& & F139M & 1384 & 21 & 3$\times$88.4 & & \\
& & F153M & 1532 & 21 & 3$\times$66.4 & & \\
\enddata
\vspace{4mm}
\textbf{Notes.}
\tablenotetext{a}{The position angle of the $y$-axis of the detector on the sky. The position angle of north on the detector is $\theta_{\rm N} = -{\tt ORIENTAT}$.}
\tablenotetext{b}{The plate scale within the FITS header are equivalent in the $x$ and $y$ directions ($p_x=p_y$).}
\tablenotetext{c}{Last image at each orientation was 3$\times$66.4\,s.}
\tablenotetext{d}{Last image at each orientation was 60.0\,s.}
\end{deluxetable*}
Four epochs of HST observations of HD 106906 were used in this study; one epoch each using the Advanced Camera for Surveys (ACS) and the Space Telescope Imaging Spectrograph (STIS), and two epochs using the Wide Field Camera 3 (WFC3). A complete observing log is given in Table~\ref{tbl:log} where the orientation of the detector on the sky and the instrument plate scale are derived from the {\tt ORIENTAT} value and {\tt CD} matrix within the FITS header.

\begin{figure}
\includegraphics[width=1.0\columnwidth]{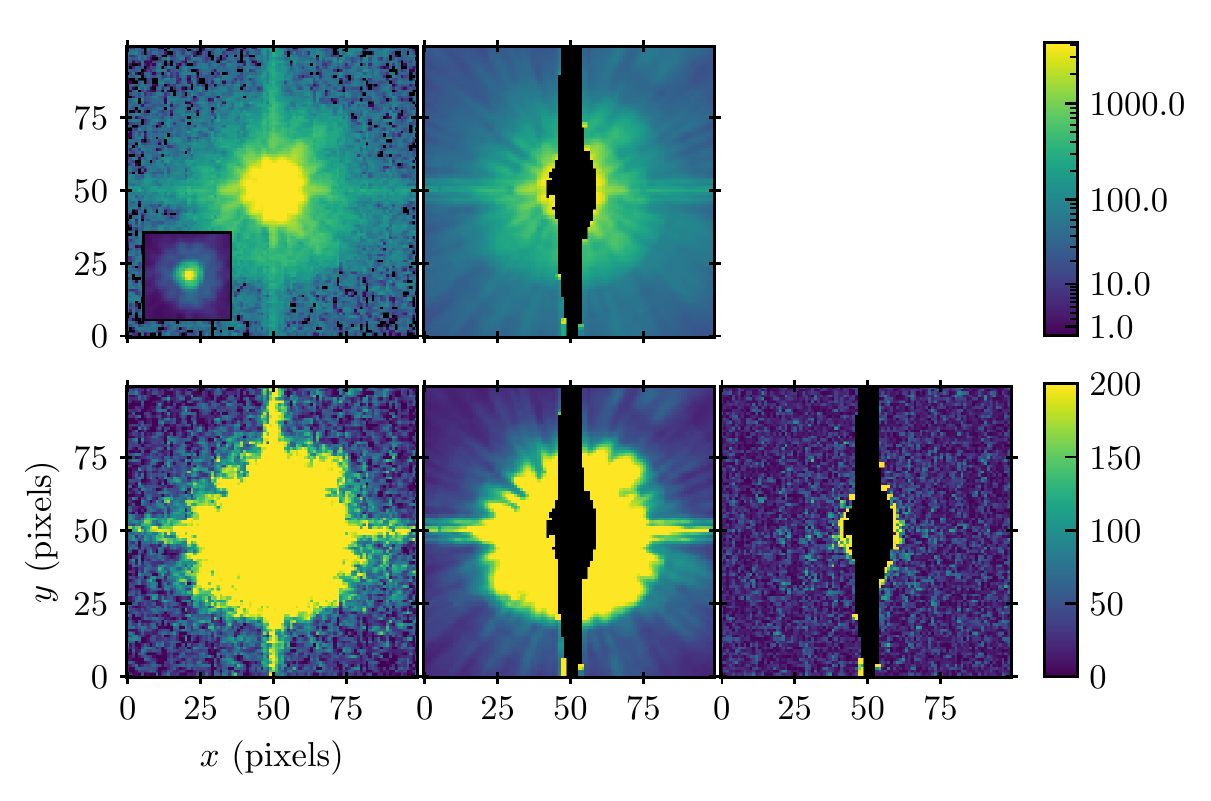}
\caption{ACS acquisition image of HD 106906 with total exposure times of 0.2082s (left column, image j91711011) and 70s (middle column, image j91711021) taken before the star was moved behind the coronagraphic spot. The inset in the top left panel shows the core of the point spread function (PSF) on a logarithmic scale between the minimum and maximum pixel value. The two images were taken consecutively with no telescope offset. A significant number of saturated pixels are present in the longer exposure (black). The absolute difference between the two images (without translation or scaling) shows no significant residuals beyond the saturated region (bottom right panel).\label{fig:acs1}}
\end{figure}
\begin{figure}
\includegraphics[width=1.0\columnwidth]{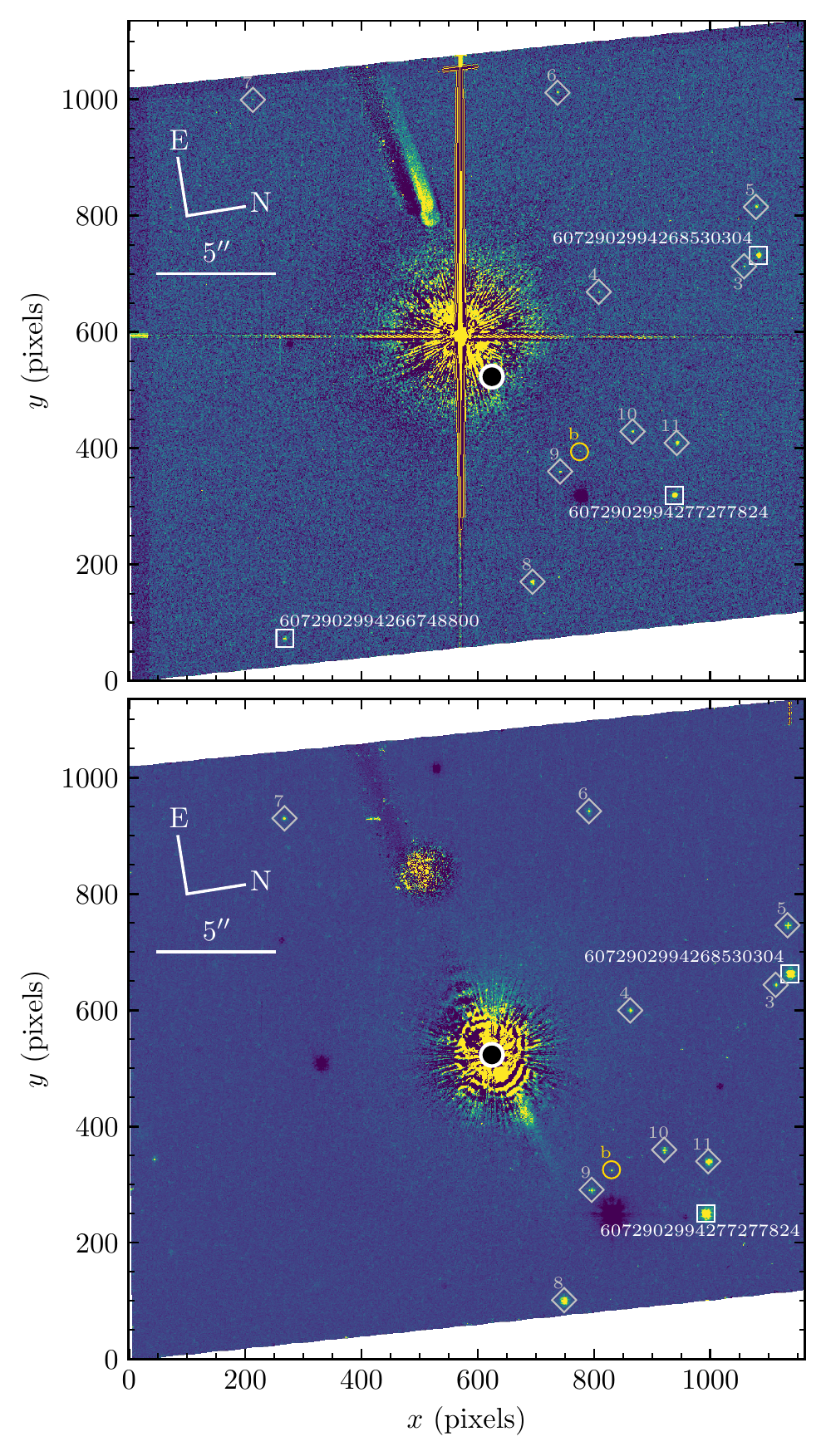}
\caption{ACS observations of HD 106906 during the acquisition procedure (top, image j91711021) and after the star had been positioned behind the coronographic spot (bottom, image j91711031). The position of the coronagraphic spot is indicated (black circle). Background stars within the Gaia DR2 catalog are indicated (white squares), as well as the additional sources used in our alignment procedure that were below the sensitivity limit of the Gaia DR2 catalog (gray diamonds). The location of HD 106906 b is indicated with a yellow circle, but is only detected at a high significance in the coronagraphic image.\label{fig:acs2}}
\end{figure}
Three sets of images of HD 106906 were taken with ACS on 2004 December 01 (GO-10330, PI Ford) using the coronagraphic mode of the high-resolution channel (HRC) with the F606W filter. The first two sets of exposures (2 $\times$ 0.1041\,s and 2 $\times$ 35\,s) were acquisition images taken before the star was placed behind the coronagraph (Figure~\ref{fig:acs1}). The star was not saturated in the first image, but heavily saturated in the second. The third exposure set (2 $\times$ 1250\,s) was taken after a small offset to position the star behind the coronagraphic mask (Figure~\ref{fig:acs2}). An image of a PSF reference star (HR 4570, F5V, $V=6.3$) was obtained on the following orbit with the same instrument configuration except for a slightly reduced exposure time (2 $\times$ 1175\,s).

\begin{figure}
\includegraphics[width=1.0\columnwidth]{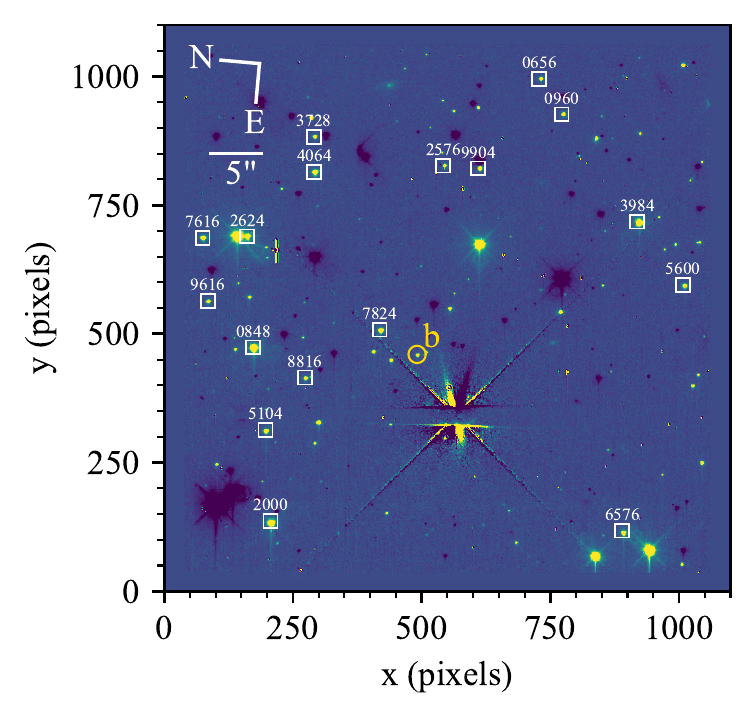}
\caption{Example STIS (od9t01010) image of HD 106906. Background sources from the Gaia DR2 catalog are indicated (white squares) and uniquely identified using the last four digits of their catalog ID number. The position of HD 106906 b is indicated with a yellow circle.\label{fig:stis_epoch}}
\end{figure}
Nine images of HD 106906 were taken with STIS on 2017 February 24 (GO-14670, PI Kalas) using the 50CORON aperture and the ``Clear'' filter (Figure~\ref{fig:stis_epoch}). The star was observed at three different telescope orientations at which three images were obtained; two 350\,s exposures near the $1.75\arcsec$ width position and one 60\,s exposure near the $1.0\arcsec$ width position, both on occulter wedge B that was used to block the starlight. The same PSF reference star was observed between the first and second telescope orientations at the same positions on the occulting wedge but with shorter exposure times (100 and 12\,s) given the relative brightness of the two stars.

\begin{figure}
\includegraphics[width=1.0\columnwidth]{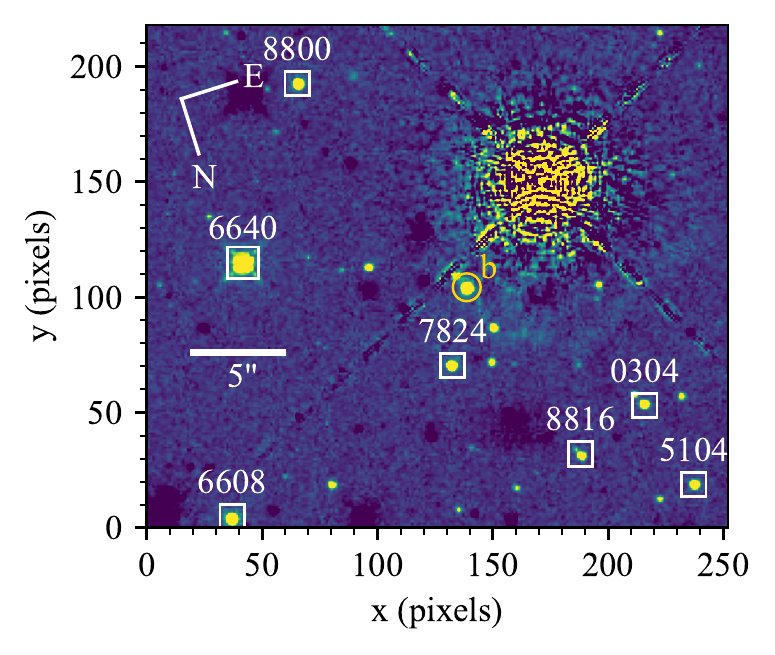}
\caption{Example 2016 WFC3 (icytb0011) image of HD 106906. Background sources from the Gaia DR2 catalog are indicated (white squares) and uniquely identified using the last four digits of their catalog ID number. The position of HD 106906 b is indicated with a yellow circle.\label{fig:wfc3_2016_sample}}
\end{figure}
HD 106906 has been observed twice with WFC3; on 2016 January 29 and 2018 June 7 (both under GO-14241, PI Apai; Figure~\ref{fig:wfc3_2016_sample}). The instrument configuration and observing strategy was similar for both epochs. The observations were taken using the IRSUB256 aperture. At both epochs groups of 10 image sets were taken at each of two orientations, cycling through each filter (f127m, f139m, f153m) with the tenth image set being an additional f127m observation. For the 2016 epoch only 20 image sets were taken, while in 2018 the cycle was repeated three times resulting in 60 image sets, 30 per telescope orientation. Integration times were the same for both epochs, 2$\times$66.4\,s, 3$\times$88.4\,s, and 3$\times$66.4\,s for the three filters, except for the last f127m image set in each cycle, which was 3$\times$66.4\,s.

\subsection{Reduction and PSF subtraction}
\label{sec:psf_subtraction}
All of the reduced observations were obtained from the Mikulski Archive for Space Telescopes.\footnote{\url{ https://archive.stsci.edu/hst/}} The data had been processed automatically by the observatory pipeline, including the typical corrections for the dark current, flat field, and, most crucially, the geometric distortion to account for anamorphism in the detector of each instrument. For the ACS and WFC3 frames we used the combined ``drizzled'' image (``\_drz''), whereas for STIS we used the individual reduced frames after the distortion correction had been applied (``\_sx2'').

We performed additional postprocessing to subtract the PSF of HD 106906. For the ACS observations we used the PSF of the reference star HR 4570 as a model for the PSF of HD 106906. Although this reference star was also observed with STIS, a background star near HR 4570 was in close proximity to HD 106906 b in several images when the two datasets were aligned, potentially biasing the companion astrometry. We instead used observations of HD 106906 taken at different telescope orientations as a reference PSF. This was also done for the WFC3 observations as no reference star was observed. For each image we optimized the PSF subtraction by shifting and scaling the reference PSF to minimize the residuals within an annulus surrounding HD 106906. The inner and outer radii were set at 50 and 100 pixels for ACS, 30 and 80 pixels for STIS (excluding regions of the annulus containing the occulting wedge or the bright diffraction spikes of the star), and 15 and 55 pixels for WFC3.

\section{Point-source astrometry}
\label{sec:astrometry}
HD 106906 is either heavily saturated or occulted by a coronagraphic optical element in all of our images within which HD 106906 b is detected at a high significance. Measuring the precise location of saturated or obscured stars is challenging, typically relying on fitting a model to the saturated PSF (e.g., \citealp{Vigan:2012jm}) or by measuring the path of the diffraction spikes (e.g., \citealp{Pueyo:2015cx}). Instead of using these approaches, we used the the Gaia Data Release 2 catalog (Gaia DR2; \citealp{GaiaCollaboration:2018io}) sources that were visible within each of our images to determine the pixel position of HD 106906. This technique is independent of assumptions made regarding the behavior of the instrumental PSF at large angular separations when fitting either the wings of a saturated star or the path of the diffraction spikes. Instead, it is limited by the ability to measure the position of unsaturated stars and the precision of the Gaia astrometric catalog.

\subsection{Fitting Gaia sources}
We queried the Gaia DR2 catalog for all sources within $120\arcsec$ of HD 106906 that had measurements of their position ($\alpha$, $\delta$), proper motion\footnote{We use the notation $\alpha^\star = \alpha \cos\delta$.} ($\mu_{\alpha^{\star}}$, $\mu_{\delta}$), and parallax ($\pi$), including HD 106906 itself. We applied a small correction to the astrometry of HD 106906 necessary due to the measured offset between the ICRS and the Gaia DR2 bright star reference frames \citep{Lindegren:2019iv,Lindegren:2020ik}, and a small correction to the astrometric uncertainties \citep{Arenou:2018dp}. We excluded Gaia sources that either fell outside of the instrument field of view, were obscured by the coronagraphic optical elements, were contaminated by the diffraction spikes of HD 106906, or were too close to a bad pixel that could not be reliably corrected. Sources with count rates exceeding the linearity limit were also excluded. One source (Gaia DR2 6072902994265468544) was rejected for being a close binary that may have biased the fitting procedure.

After applying these cuts we were left with either two or three useful sources for ACS, between 15 and 21 useful sources for STIS, and between five and seven useful sources for WFC3. For each background star, we extracted an 11-by-11 pixel (approximately $5\lambda/D$ for ACS and $10\lambda/D$ for WFC3 and STIS) data stamp centered on the predicted position of the star and iteratively fit a 2D symmetric Gaussian using a Levenberg--Marquardt least-squares fitting algorithm. A constant offset was not fit as the background was subtracted by the PSF subtraction procedure described previously. We used the same algorithm to measure the coordinates of HD 106906 b within each image, the additional sources not within Gaia used to align the ACS images (see Section~\ref{sec:astro_model}), and HD 106906 in the first acquisition image taken with ACS where the star is not saturated (see Section~\ref{sec:application}). The measured pixel positions for each background star in each image, and their propagated tangent plane offsets relative to HD 106906 using the model described in Section~\ref{sec:astro_model}, are given in Table~\ref{table:dr2_catalogue}.

\subsection{Astrometric uncertainties}
\label{sec:errors}
\begin{figure}
\includegraphics[width=1.0\columnwidth]{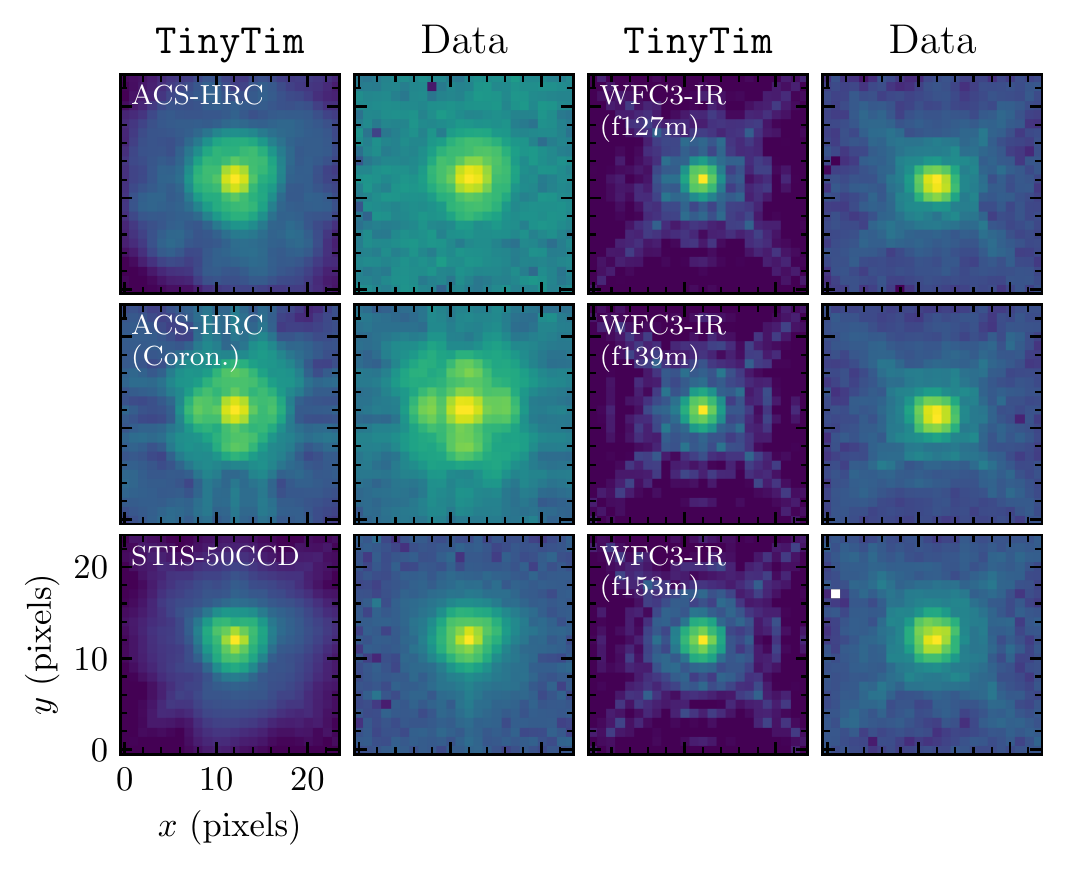}
\caption{\texttt{TinyTim} PSFs (first and third columns) and representative examples extracted from the ACS, STIS, and WFC3 observations (second and fourth columns). The simulated PSFs are noiseless and were created assuming no telescope jitter or defocus. \label{fig:error1}}
\end{figure}
\begin{figure}
\includegraphics[width=1.0\columnwidth]{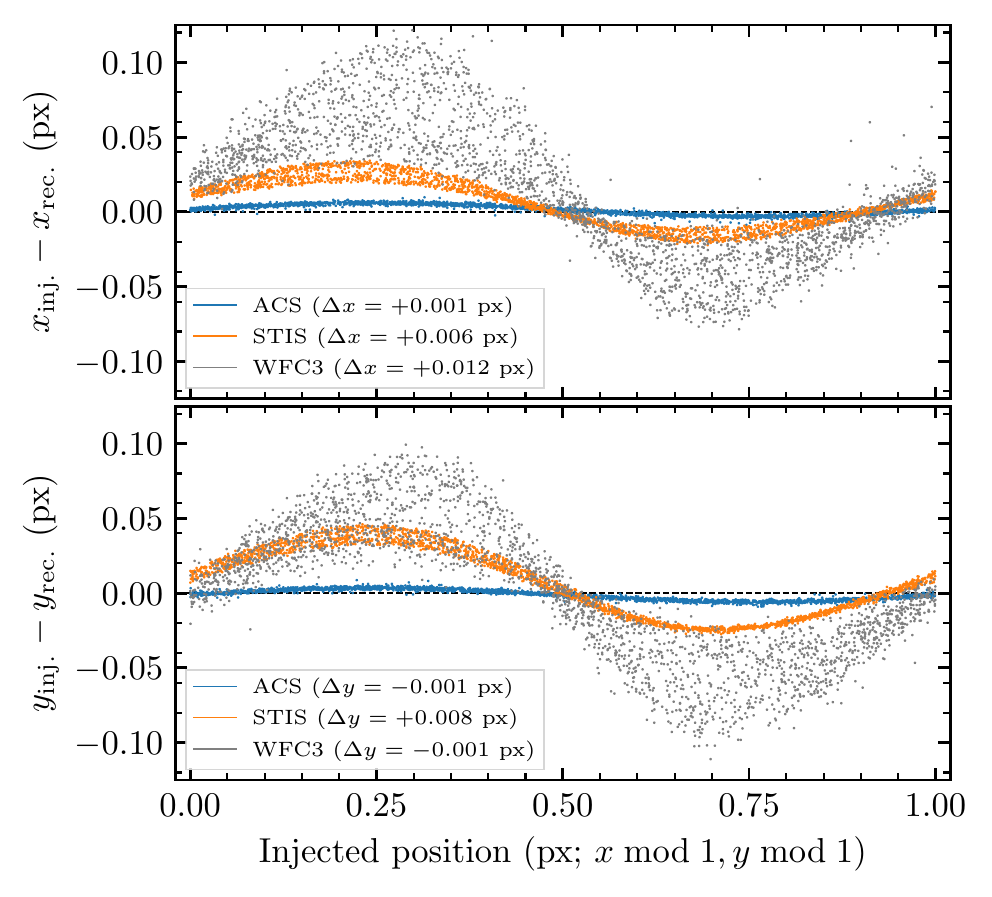}
\caption{Difference between injected and recovered position in $x$ (top) and $y$ (bottom) as a function of the subpixel position of the injected PSF ($x\bmod 1$, $y\bmod 1$) for one source within one image of the ACS (blue), STIS (orange), and WFC3 (f153m, gray) datasets. The median offset for each instrument is given in the figure legend.\label{fig:error2}}
\end{figure}

We used an injection and recovery framework to assess the uncertainty of the measurements of each background star and HD 106906 b in each image. We used \texttt{TinyTim} \citep{Krist:2011kt} to generate synthetic PSFs for each instrument configuration. The objective here was not to perfectly match the PSF of the stars within each image, rather it was to use a reasonably good approximation of the PSF to estimate the precision and accuracy of our Gaussian model. Oversampled PSFs were generated using the spectrum of a K7V star without any focus despace or jitter applied. The resulting PSFs were a reasonable approximation of the observations (Figure~\ref{fig:error1}).

Five thousand injections were performed per source at random locations within an annulus surrounding the source. The width of the annulus was five pixels, with a central radius of 20 pixels for ACS and STIS and 15 pixels for WFC3. The oversampled PSF was shifted by cubic interpolation to the subpixel position of the injection, pixelated to match the sampling of the detector, normalized to have a flux within 10\% of the source being tested, and added into the image. The injected PSF was fit using the Gaussian model described previously.

This analysis revealed three sources of error; a random error associated with the signal-to-noise ratio of the injected source and two systematic errors, one associated with the asymmetry of the PSF and the other with the subpixel position of the injected source. The asymmetry of the \texttt{TinyTim} PSF was accounted for by adding a small correction to the recovered pixel position, estimated from a Gaussian fit to a 10 times oversampled \texttt{TinyTim} PSF generated for each instrument configuration. This offset ($\Delta x = x_{\rm injected} - x_{\rm recovered}$) was estimated to be $\Delta x=-0.008$\,px and $\Delta y=-0.012$\,px for the unocculted ACS PSF, $\Delta x=-0.006$\,px and $\Delta y=-0.031$\,px for the occulted ACS PSF, $\Delta x=-0.054$\,px and $\Delta y=-0.047$\,px for the STIS PSF, and $\Delta x=-0.013$\,px and $\Delta y=0.0$\,px for the WFC3/IR PSF. The systematic error associated with the subpixel position of the injected source is demonstrated in Figure~\ref{fig:error2}, after applying the correction due to the asymmetry of the instrument PSF. This effect was not seen at a significant level for the ACS PSF as this instrument is not undersampled. These results were not sensitive to the choice of spectral type used to construct the \texttt{TinyTim} PSF. We repeated the experiment for spectral types between K4 and M3, a probable range of spectral types for the distant background objects based on the small subset with effective temperatures reported in the Gaia catalogue, and the values of the derived systematic and random errors were not significantly different.

We accounted for the random error attributed to the signal-to-noise ratio and this remaining systematic error by conservatively adopting the 95th percentile of the absolute value of the difference between the injected and recovered position as our 1$\sigma$ uncertainty. These uncertainties ($\sigma_x$, $\sigma_y$) were calculated separately in the $x$ and $y$ directions. Our preliminary fits of the positions of the STIS background stars yielded $\chi^2_{\nu}\sim~2$, indicating that our uncertainties were slightly underestimated. We combined in quadrature the uncertainties estimated for each source with an additional uncertainty of 0.05\,px, leading to $\chi^2_{\nu}\sim~1$ for these data.

\section{Astrometric model}
\label{sec:astro_model}
\subsection{Epoch propagation}
We propagated the astrometric measurements for each source from the Gaia reference epoch (2015.5) to the epoch of the observation using a Monte Carlo approach. We started by generating $10^5$ random draws for each of the five astrometric parameters from a multivariate Gaussian distribution created from the catalog values, uncertainties, and covariances. These random draws were then propagated from 2015.5 to the appropriate epoch using a rigorous coordinate transformation accounting for perspective effects \citep{Butkevich:2014jt}. We adopted a radial velocity of $12.18\pm0.15$\,km\,s$^{-1}$ for HD 106906, while assuming zero radial velocity for the background stars. These stars are likely so distant that their radial velocity has negligible effect on the coordinate propagation. The propagated spherical coordinates of each source ($\alpha$, $\delta$) were then transformed into tangent plane offsets relative to the propagated position of HD 106906 ($\alpha_0$, $\delta_0$) following \citet{Bedin:2018kz} as
\begin{equation}
\begin{aligned}
    \xi_{\rm sky} = & \frac{\cos\delta\sin(\alpha - \alpha_0)}{\sin\delta_0\sin\delta + \cos\delta_0\cos\delta\cos(\alpha - \alpha_0)} \\ 
    & + \left(\pi P_\alpha - \pi_0 P_{\alpha, 0}\right),\\
    \eta_{\rm sky} = & \frac{\cos\delta_0 \sin\delta - \sin\delta_0  \cos\delta \cos(\alpha - \alpha_0)}{\sin\delta_0 \sin\delta + \cos\delta_0 \cos\delta \cos(\alpha - \alpha_0)} \\ 
    & + \left(\pi P_\delta - \pi_0 P_{\delta, 0}\right).
\end{aligned}
\end{equation}
The additional terms for both $\xi_{\rm sky}$ and $\eta_{\rm sky}$ account for the nonzero parallax of both HD 106906 ($\pi_0$) and the background stars ($\pi$). The parallax factors in the $\alpha^\star$ and $\delta$ directions ($P_{\alpha,0}$, $P_{\delta,0}$ for HD 106906 and $P_\alpha$, $P_\delta$ for each source) were calculated from the Euclidean coordinates of the Earth relative to the solar system barycenter ($X_\oplus$, $Y_\oplus$, $Z_\oplus$) at the epoch of the observation,
\begin{equation}
\begin{split}
    P_{\alpha} & = X_\oplus \sin\alpha - Y_\oplus \cos\alpha,\\
    P_{\delta} & = X_\oplus \cos\alpha \sin\delta +
Y_\oplus \sin\alpha \sin\delta - Z_\oplus \cos\delta.
\end{split}
\end{equation}
This process resulted in $10^5$ values for both $\xi_{\rm sky}$ and $\eta_{\rm sky}$ for each of the background stars at each epoch. We adopted the mean of these draws as the tangent plane offset, and calculated the covariance between them to account for correlated uncertainties within the Gaia catalog. The tangent plane offset and associated correlation coefficient for each source at each epoch is given in Table~\ref{table:dr2_catalogue}.

\subsection{Converting to detector coordinates}
The tangent plane offsets ($\xi_{\rm sky}$, $\eta_{\rm sky}$) were converted to detector coordinates ($\xi$, $\eta$) using the five parameters within our model---the pixel position of HD 106906 ($x_0$, $y_0$), and the pixel scale ($p_x$, $p_y$) and the position angle of north ($\theta_{\rm N}$) on the detector---as
\begin{equation}
\begin{bmatrix}\xi \\ \eta\end{bmatrix} = \left(\mathbf{R}\begin{bmatrix}\xi_{\rm sky} \\ \eta_{\rm sky}\end{bmatrix}\right) \begin{bmatrix}1/p_x \\ 1/p_y\end{bmatrix} + \begin{bmatrix}x_0 \\ y_0 \end{bmatrix},
\end{equation}
and $\mathbf{R}$ the rotation matrix
\begin{equation}
\mathbf{R}  = \begin{bmatrix}-\cos\theta_{\rm N} & \sin\theta_{\rm N} \\ \sin\theta_{\rm N} & \cos\theta_{\rm N}\end{bmatrix},
\end{equation}
to rotate and scale the tangent plane offsets to match the orientation and pixel scale of the detector. The rotation matrix has a nonstandard form to account for the flip in the $x$-axis when transforming between detector and sky coordinates. We also rotated and scaled the covariance matrix
\begin{equation}
    \mathbf{C}_{\xi\eta} = \left(\mathbf{R} \mathbf{C_{\xi\eta, {\rm sky}}} \mathbf{R}^T\right) \circ \mathbf{S},
\end{equation}
where $\circ$ represents an element-wise multiplication. The scaling matrix $\mathbf{S}$ is defined as 
\begin{equation}
\mathbf{S}  = \begin{bmatrix}1/p_x^2 & 1/p_xp_y \\ 1/p_xp_y & 1/p_y^2\end{bmatrix}.
\end{equation}

\subsection{Goodness of fit}
The predicted position of a background star ($\xi$, $\eta$) was compared to the measured pixel position ($x$, $y$) and adopted uncertainty ($\sigma_x$, $\sigma_y$) to determine a goodness of fit $\chi^2$ for a given set of model parameters as
\begin{equation}
    \chi^2 = \vec{r}^T \mathbf{C}^{-1} \vec{r}
\end{equation}
where $\vec{r}$ is the residual vector
\begin{equation}
    \vec{r} = \begin{bmatrix}\xi - x \\ \eta - y\end{bmatrix},
\end{equation}
and $\mathbf{C}$ is the covariance matrix created from the combination of $\mathbf{C_{\xi\eta}}$ and a covariance matrix describing the uncertainty on the measured pixel position of the background star, assumed to be completely uncorrelated
\begin{equation}
    \mathbf{C} = \mathbf{C_{\xi\eta}} + \begin{bmatrix}\sigma_x^2 & 0 \\ 0 & \sigma_y^2\end{bmatrix}.
\end{equation}
For the STIS measurements, $\sigma_x$ and $\sigma_y$ here are the quadratic sum of the measurement uncertainty and the 0.05\,px error inflation term described in Section~\ref{sec:errors}.

\subsection{Specifics for ACS}
There are only two Gaia sources within the coronagraphic ACS image, severely limiting our ability to constrain the location of HD 106906. Instead, we used the three Gaia sources within the second acquisition image to determine the star location in this image ($x_0^{\rm acq}$, $y_0^{\rm acq}$) and measured the offset between the acquisition and coronagraphic image using the 11 sources common to both images, yielding the location of the star in the coronagraphic image  ($x_0^{\rm coro}$, $y_0^{\rm coro}$). The star position in the acquisition image was fit using the procedure described previously. The offset between the two images was determined by adding three additional parameters to our model; a counterclockwise rotation ($\Delta \theta$) about $(0,0)$, and a subsequent translation ($\Delta x$, $\Delta y$) to align the 11 stars common to both images. The goodness of fit for this alignment step was calculated as
\begin{equation}
    \chi^2_{\rm align} = \vec{r}^T\mathbf{C}^{-1}\vec{r}
\end{equation}
where
\begin{equation}
\begin{split}
    \vec{r} &= \mathbf{R}\begin{bmatrix}x^{\rm acq} \\ y^{\rm acq} \end{bmatrix}+\begin{bmatrix}\Delta x-x^{\rm coro} \\  \Delta y -y^{\rm coro} \end{bmatrix},\\
    \mathbf{C} &= \mathbf{R}\mathbf{C}^{\rm acq}\mathbf{R}^T + \mathbf{C}^{\rm coro},\\
    \mathbf{R} & = \begin{bmatrix}\cos\Delta\theta & -\sin\Delta\theta \\ \sin\Delta\theta & \cos\Delta\theta\end{bmatrix}.
\end{split}
\end{equation}
where $x^{\rm acq}$, $y^{\rm acq}$ and $x^{\rm coro}$, $y^{\rm coro}$ are the pixel position of the background star in the acquisition and coronagraphic image, and $\mathbf{C}^{\rm acq}$ and $\mathbf{C}^{\rm coro}$ are the corresponding covariance matrices. As previously, the measurements were assumed to be uncorrelated and these were used here for completeness. The goodness of fit for a given set of parameters was calculated from the $\chi^2$ of the fit of the Gaia sources in the acquisition image and $\chi^2_{\rm align}$ as
\begin{equation}
    \chi^2 = \sum_i^{n_{\rm gaia}}\chi^2_{{\rm Gaia},i} + \sum_j^{n_{\rm align}}\chi^2_{{\rm align},j}
\end{equation}
for the $n_{\rm gaia}=3$ Gaia sources in the acquisition image and the $n_{\rm align}=11$ sources common to both images. With either the best-fit parameters, or their posterior distributions, the pixel position of HD 106906 in the coronagraphic image can be calculated as
\begin{equation}
    \begin{bmatrix}x_0^{\rm coro} \\ y_0^{\rm coro}\end{bmatrix} = \begin{bmatrix}\cos\Delta\theta & -\sin\Delta\theta \\ \sin\Delta\theta & \cos\Delta\theta\end{bmatrix} \begin{bmatrix}x_0^{\rm acq}  \\ y_0^{\rm acq}\end{bmatrix} + \begin{bmatrix}\Delta x \\  \Delta y \end{bmatrix}.
\end{equation}

\section{Application to HD 106906}
\label{sec:application}
We considered four variants of the model described in Section~\ref{sec:astro_model} to determine the position of HD 106906 in each of our images. The first only had the coordinates of HD 106906 ($x_0$, $y_0$) as free parameters, using the pixel scale and detector orientation given within the FITS header. The remaining three variants had various combinations of the position angle of north on the detector ($\theta_{\rm N}$) and the pixel scales along the $x$- and $y$-axes of the detector ($p_x$, $p_y$) as free parameters to explore the effects of varying the instrument calibration values (model 1: $x_0y_0$, model 2: $x_0y_0\theta_{\rm N}$, model 3: $x_0y_0p_xp_y$, model 4: $x_0y_0\theta_{\rm N} p_xp_y$). 

As described in Section~\ref{sec:astro_model}, only two Gaia sources were present within the deep ACS observation in which the planet was visible due to the small field of view of the instrument (Figure~\ref{fig:acs2}, top panel). A third source was visible when the star was offset slightly from the coronagraphic mask for the acquisition images (Figure~\ref{fig:acs2}, bottom panel). To include this third star in our analysis we modified the model to simultaneously fit both the position of HD 106906 in the acquisition image and the translation and rotation between the acquisition and coronagraphic images using the two Gaia sources and nine additional sources not within the Gaia catalog that are common to both images.

\subsection{Model selection and parameter estimation}
We used a maximum likelihood estimation to find the best fit parameters for each of the four variants of our model when applied to each image. The complexity of the model adopted for each epoch was justified by comparing the average value of the Bayesian information criterion (BIC) for each variant of the model (Fig~\ref{fig:BIC}). The BIC is a metric that reduces overfitting by penalizing the excessive use of free parameters in a model. For both ACS and WFC3 we found small values of $\Delta$BIC ranging from 0.5 to 6.6 when comparing the more complex models to the two-parameter model, suggesting that the improvement in $\chi^2$ was not sufficient to justify the additional parameters. For the STIS observations we found a large $\Delta$BIC of 1352 when comparing the two-parameter and five-parameter models, indicating that the significant improvement in the goodness of fit justifies the use of the additional parameters for this model. This large improvement in the quality of the fit was due to a systematic error of the position angle of north reported in the FITS header for the STIS observations (incorrect by $0\fdg077$ for these data; see Appendix~\ref{sec:hst-calib}).

We used the affine-invariant Markov Chain Monte Carlo ensemble sampler \texttt{emcee} \citep{ForemanMackey:2013io} to sample the posterior distributions of the fitted parameters. We used uniform bounded priors on all free parameters, excluding nonphysical plate scales and limiting the orientation between 0 and $2\pi$. We initialized 50 walkers near the maximum likelihood and advanced them for 1500 steps. The first 500 steps of each chain was discarded as a ``burn-in'' where the walker positions are still a function of their initial positions. The values and corresponding uncertainties for the fitted parameters for each image of each epoch are given in Tables \ref{tbl:acs-results}, \ref{table:stis_5param}, \ref{table:wfc3_2016_2param}, and \ref{table:wfc3_2018_2param} of the Appendix.

\subsection{Relative astrometry}
\begin{deluxetable*}{ccccccc}
\tabletypesize{\normalsize}
\tablecaption{Relative astrometry between HD 106906 and HD 106906 b.\label{table:rel_astrometry}}
\tablehead{
\colhead{UT Date} & \colhead{MJD} & \colhead{Instrument} & \colhead{$\Delta \alpha^\star$ (arcsec)} & \colhead{$\Delta \delta$ (arcsec)} & \colhead{$\rho$ (arcsec)} & \colhead{$\theta$ (deg)}}
\startdata
2004-12-01 & 53340.7 & ACS & $-5.6903\pm0.0054$ & $4.2997\pm0.0054$ & $7.1321\pm0.0054$ & $307.076\pm0.043$\\
2016-01-29 & 57416.9 & WFC3 (f127m) & $-5.6563\pm0.0196$ & $4.3128\pm0.0159$ & $7.1130\pm0.0183$ & $307.325\pm0.140$ \\
2016-01-29 & 57416.9 & WFC3 (f139m) & $-5.6581\pm0.0172$ & $4.3152\pm0.0145$ & $7.1158\pm0.0163$ & $307.331\pm0.125$\\
2016-01-29 & 57416.9 & WFC3 (f153m) & $-5.6539\pm0.0127$ & $4.3096\pm0.0114$ & $7.1091\pm0.0123$ & $307.316\pm0.096$\\
2017-02-24 & 57808.8 & STIS & $-5.6585\pm0.0039$ & $4.3058\pm0.0037$ & $7.1104\pm0.0038$ & $307.270\pm0.030$\\
2018-06-07 & 58276.1 & WFC3 (f127m) & $-5.6565\pm0.0175$ & $4.3086\pm0.0179$ & $7.1106\pm0.0176$ & $307.296\pm0.143$\\
2018-06-07 & 58276.1 & WFC3 (f139m) & $-5.6573\pm0.0157$ & $4.3091\pm0.0159$ & $7.1115\pm0.0158$ & $307.296\pm0.128$\\
2018-06-07 & 58276.1 & WFC3 (f153m) & $-5.6568\pm0.0120$ & $4.3126\pm0.0120$ & $7.1133\pm0.0120$ & $307.321\pm0.097$\\
\enddata
\end{deluxetable*}
\begin{figure}
\includegraphics[width=\columnwidth]{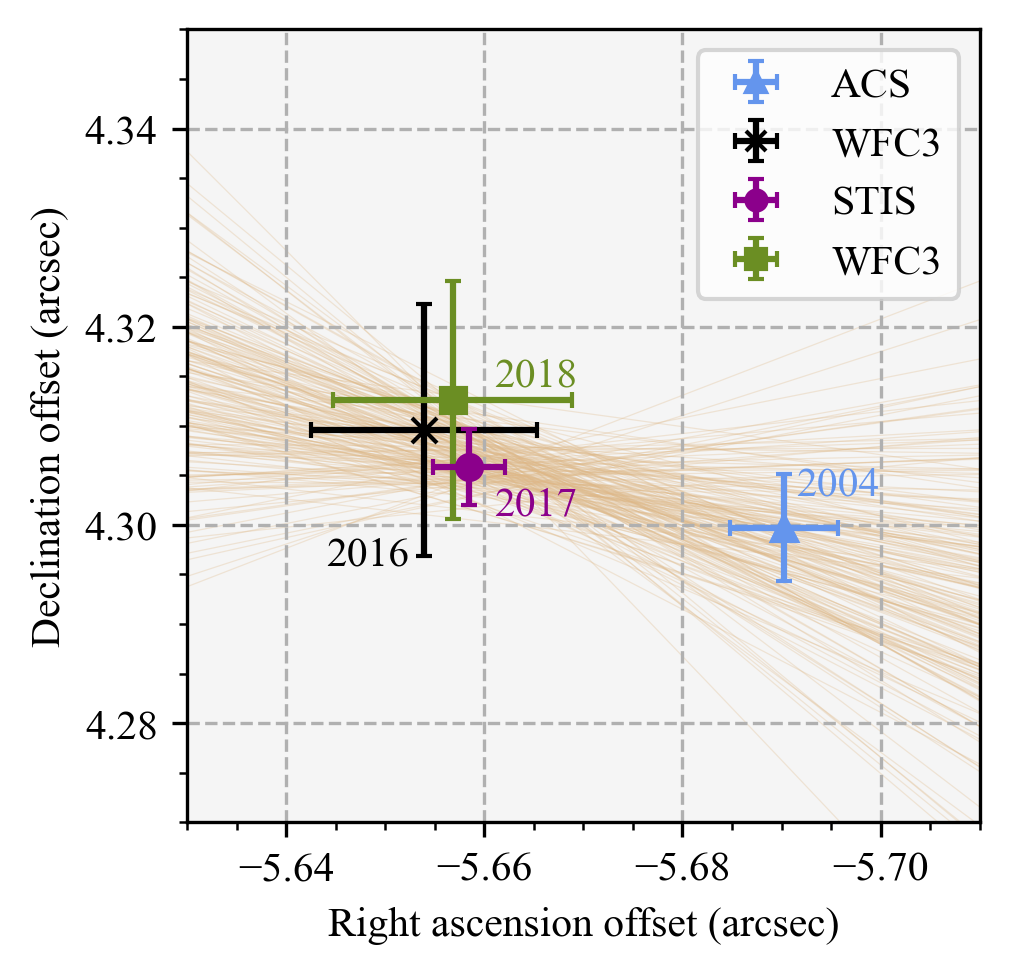}
\caption{Measured relative astrometry for HD 106906 b using ACS (blue triangle, 2004), STIS (red circle, 2017), and the f153m filter for WFC3 (black cross, 2016; green square, 2018; the f127m and f139m filters were excluded from this figure for visual clarity). Sample orbit fit tracks drawn from the orbital parameter posteriors are overplotted in brown.\label{fig:positions}}
\end{figure}

Relative astrometry between the star and planet was calculated by combining the posterior distributions for the pixel coordinate of the star with the measured pixel position and uncertainty of the planet. This was converted into a sky separation and position angle using the detector plate scales and orientation. The final astrometry for each epoch (and filter for WFC3) was calculated using a weighted mean of the measurements ($x_i\pm\sigma_i$) from each image. For each offset ($\Delta\alpha^\star$, $\Delta\delta$) the weighted mean $\bar{x}$ and corresponding uncertainty $\bar{\sigma}$ were calculated as $\bar{x}=\sum_iw_ix_i/\sum_iw_i$ and $\bar{\sigma}=\sqrt{\sum_i\sigma_i^2w_i/\sum_iw_i}$ where $w_i=1/\sigma_i^2$, under the assumption that the measurements were not independent.

For the ACS observations we used the measurement of the star position in the first acquisition image as a prior for the location in the second under the assumption that there was no telescope motion between these images (Figure~\ref{fig:acs1}). We calculated a maximum offset between the two acquisition images of 0.2\,px; larger offsets produced significant residuals when the two images were subtracted. We fit the ACS data with and without this prior and find consistent relative astrometry (Table~\ref{tbl:acs-results}), and we adopted the measurements derived from the fit using this prior as our final relative astrometry for the ACS epoch. The final astrometric measurements from each epoch are given in Table~\ref{table:rel_astrometry}, and plotted in Figure~\ref{fig:positions}. Our relative astrometric measurements derived from the ACS and WFC3 observations are consistent with previous measurements \citep{Bailey:2014et,Zhou:2020iy}, with reduced uncertainties.

\subsection{Testing for Gaia systematics}
\label{sec:systematics}
\begin{figure}
\includegraphics[width=\columnwidth]{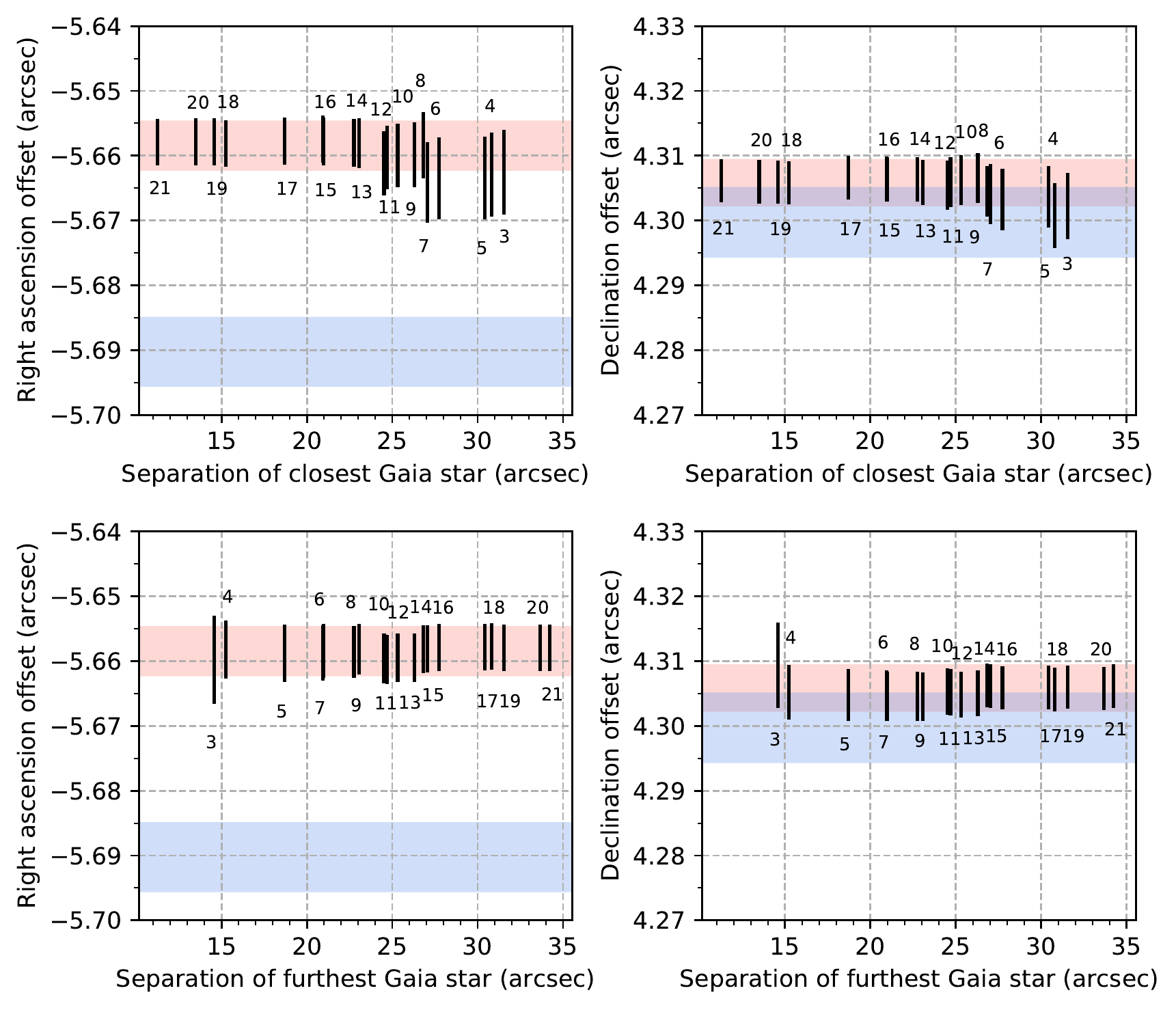}
\caption{Measured companion astrometry for STIS image od9t03020, where the number of Gaia sources included in the measurement is being iteratively changed by removing either the closest source from HD 106906 (top row) or the furthest source from HD 106906 (bottom row). The numeric labels correspond to the number of Gaia sources included in the data point next to it. The final measured position and 1$\sigma$ uncertainty for STIS (highlighted in red) and ACS (highlighted in blue) are shown for comparison.\label{fig:exclusion_radius}}
\end{figure}
We performed several experiments to gauge the consistency of our astrometry and search for potential biases, specifically testing to determine if the brightness of HD 106906 is biasing the measured Gaia astrometry of the nearby faint sources. The experiments we performed were all variations on using subsets of the available Gaia sources to obtain our final astrometry. In one experiment using the STIS frames, we limited our analysis to the same three Gaia background sources visible in the ACS observation to see if there was a small sample selection bias affecting the ACS astrometry. We did not uncover any significant bias in limiting our analysis to just these three stars. The measurement remained within 1$\sigma$ of the STIS astrometry that used all Gaia sources, albeit with an increased uncertainty due to the smaller number of stars used.

In another experiment, we iteratively fit our model for the location of HD 106906, removing the closest star with each iteration until there were only three Gaia stars left. This experiment showed no obvious systematics, with the measured positions remaining around the same value (Figure~\ref{fig:exclusion_radius}, top row). The uncertainty on each measurement scaled as a function of the decreasing number of sources used. We also tried this same experiment but in the reverse direction, iteratively fitting our model by removing the farthest star from HD 106906 with each iteration. This analysis yielded the same result, with no significant change in the relative astrometry as a function of exclusion radius (Figure~\ref{fig:exclusion_radius}, bottom row).

We also conducted experiments to test the effect of a poor PSF subtraction on the ACS epoch where the background stars used to cross-register with the Gaia catalogue are within the wings of the PSF of HD 106906, significantly closer than for the STIS epoch. We found that the final relative astrometry was insensitive to errors of the scaling of the reference star of $10$\,\%, and of relative shifts between the two stars of $1$\,pixel, both relative to the best-fit scaling factor and offsets found using the method described in Section~\ref{sec:psf_subtraction}. Errors of this magnitude produced PSF-subtracted images with significant residuals relative to the best-fit solution suggesting that the scaling and offsets were determined to a far greater accuracy.

\section{Results}
\label{sec:results}
We detected a statistically significant $31.8\pm7.0$\,mas eastward motion of the planet relative to the host star between the 2004 ACS and 2017 STIS epochs (Figure~\ref{fig:positions}). We significantly reduced the uncertainty on the ACS measurement relative to previous attempts \citep{Bailey:2014et} and made a more modest improvement over previous WFC3 astrometry \citep{Zhou:2020iy}.The main source of improvement was our usage of Gaia to cross-register background sources in order to indirectly triangulate the position of the primary star. Previous attempts tried to directly fit the PSF of the primary but were typically limited by occultation or saturation effects. The STIS measurement has not previously been published, and is consistent with the WFC3 measurements taken at a similar epoch.

\subsection{Orbital motion}
\label{sec:orbits}
\begin{deluxetable}{lccc}
\tabletypesize{\normalsize}
\tablecaption{Orbital elements and derived parameters for HD 106906 b.
\label{tbl:orbit}}
\tablehead{
\colhead{Parameter} & \colhead{Unit} & \multicolumn{2}{c}{Median ($\pm1\sigma$)}}
\startdata
Period ($P$)                           & yr & \multicolumn{2}{c}{$15000_{-6400}^{+17000}$} \\
Semi-major axis ($a$)                  & au & \multicolumn{2}{c}{$850_{-260}^{+560}$} \\
Periastron distance ($r_{\rm peri}$)   & au & \multicolumn{2}{c}{$510_{-320}^{+480}$}\\
Eccentricity ($e$)                & \nodata & \multicolumn{2}{c}{$0.44_{-0.31}^{+0.28}$}\\
Inclination ($i$)                     & deg & \multicolumn{2}{c}{$56_{-21}^{+12}$}\\
Epoch of periastron\tablenotemark{a} ($\tau$)      & \nodata & \multicolumn{2}{c}{$0.26_{-0.15}^{+0.28}$}\\
Epoch of periastron ($t_0$)            & yr & \multicolumn{2}{c}{$3200_{-1200}^{+7800}$}\\
Parallax ($\pi$)                      & mas & \multicolumn{2}{c}{$9.677\pm0.043$}\\
System mass ($M_{\rm total}$) & $M_{\odot}$ & \multicolumn{2}{c}{$2.71\pm0.14$}\\
\hline
& & Family 1\tablenotemark{b} & Family 2\tablenotemark{c} \\
\hline
PA of ascending node ($\Omega$) & deg & $99_{-28}^{+26}$ & $279_{-29}^{+25}$\\
Argument of periastron ($\omega$) & deg & $356_{-86}^{+61}$ & $176_{-87}^{+61}$\\
Mutual inclination ($i_{\rm m}$) & deg & $36_{-14}^{+27}$ & $44_{-14}^{+27}$\\
\enddata
\vspace{4mm}
\textbf{Notes.}
\tablenotetext{a}{\ $\tau = (t_0-2020) \mod P$.}
\tablenotetext{b}{\ $10\le\Omega<190$\,deg.}
\tablenotetext{c}{\ $190\le\Omega<360$\,deg or $0\le\Omega<10$\,deg.}
\end{deluxetable}
\begin{figure*}
\includegraphics[width=\linewidth]{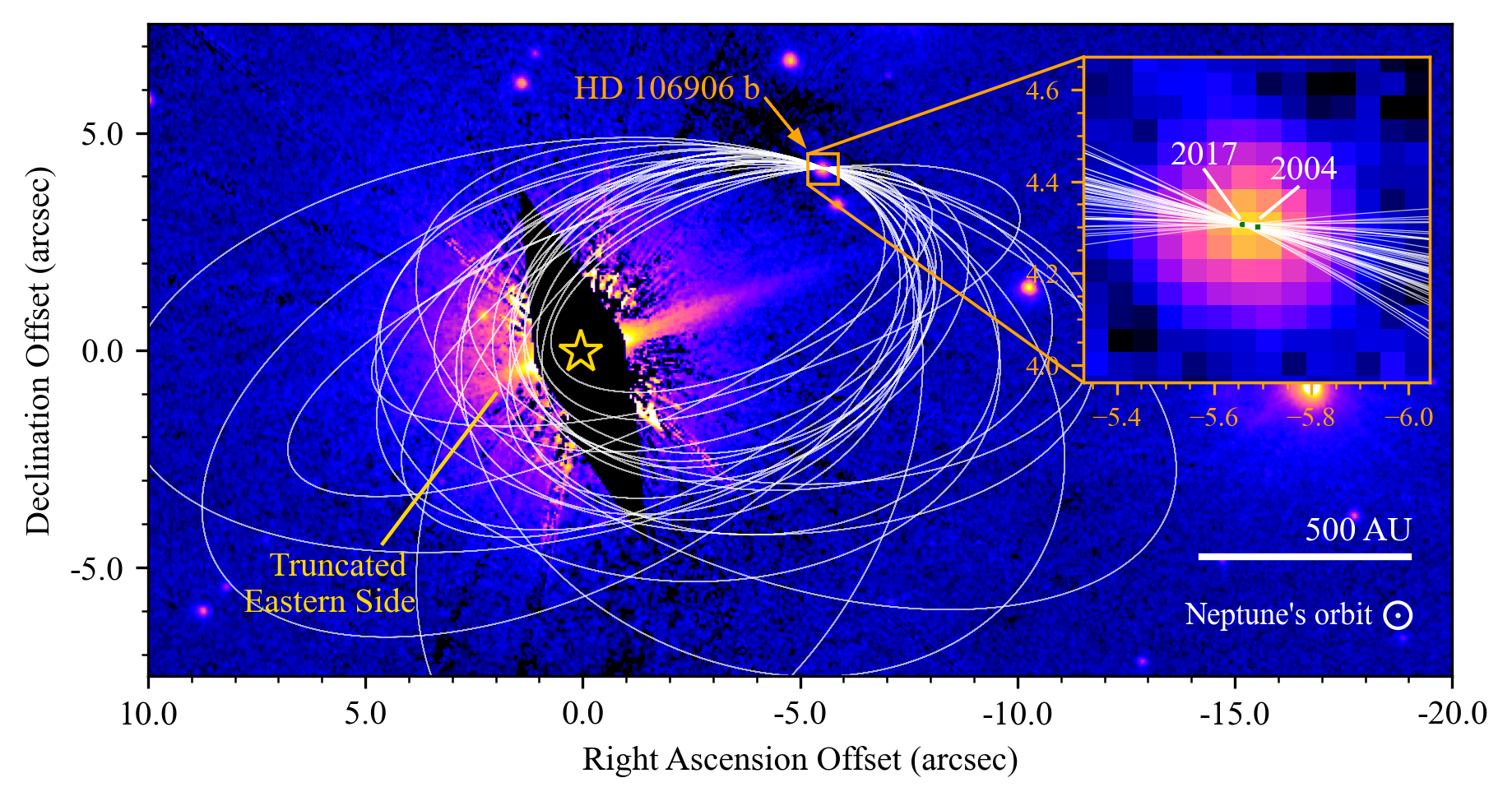}
\caption{A PSF-subtracted STIS image taken in 2017 of the HD 106906 system showing the nearly edge-on asymmetric debris disk and the planetary-mass companion HD 106906 b (top center; \citealp{Kalas:2015en}). The location of the central binary star is denoted by the star symbol. The black region near the star is a region obscured by the occulting wedge of STIS. The cross-like pattern centered on the star is a residual of the imperfect PSF subtraction. Twenty-five orbit tracks sampled randomly from the posterior distributions of the orbital elements are overplotted. The two most constraining astrometric data points from ACS and STIS are inset to highlight our measurements and to indicate the direction of the orbit of the companion. The color scale has been chosen to accentuate the disparity in the visible extension of the eastern side of the disk vs. the western side. The size of Neptune’s orbit is displayed beneath the scale bar for visual comparison.  \label{fig:orbit_tracks}}
\end{figure*}
\begin{figure*}
\includegraphics[width=\linewidth]{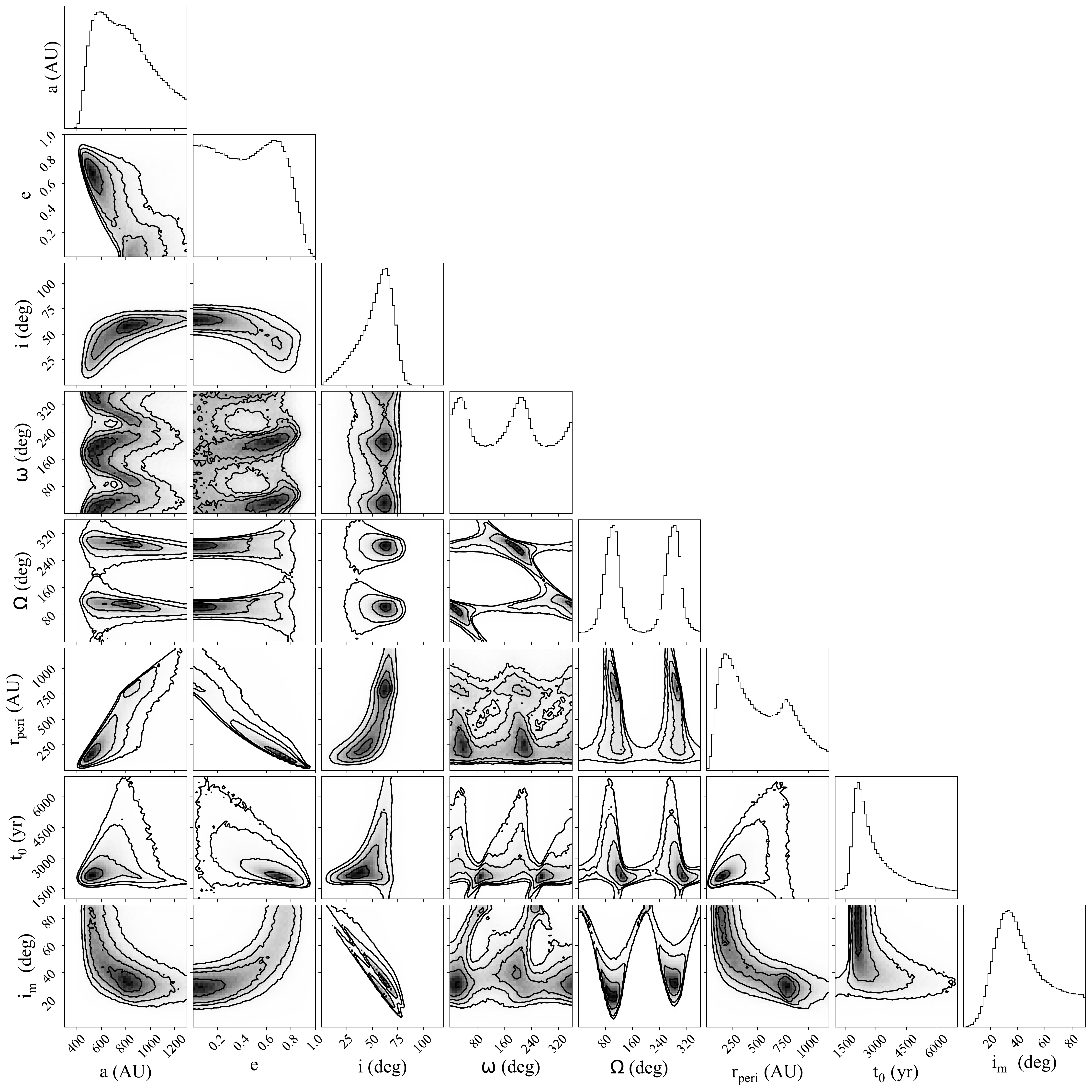}
\caption{Corner plot showing 2D covariances and 1D marginalized posterior distributions of eight orbital parameters for HD 106906 b derived using the OFTI rejection-sampling algorithm. In order of appearance, the parameters included are (1) semi-major axis, $a$; (2) eccentricity, $e$; (3) inclination, $i$; (4) argument of periastron, $\omega$; (5) longitude of the ascending node, $\Omega$; (6) periastron distance, $r_{\rm peri}$; (7)  epoch of periastron, $t_0$; and (8) mutual inclination, $i_m$.\label{fig:corner_plot_OFTI}}
\end{figure*}

We fitted the measured astrometry with a Keplerian orbit using the Orbits for the Impatient (OFTI; \citealp{Blunt:2017et}) algorithm that is part of the \texttt{orbitize} software package \citep{Blunt:2019vq}. OFTI uses a rejection-sampling algorithm to efficiently generate Bayesian posterior distributions of the orbital parameters. We fitted the following parameters: (1) semi-major axis, (2) eccentricity, (3) inclination, (4) argument of periastron, (5) position angle of the ascending node (where the companion has a positive radial velocity moving away from the observer), (6) epoch of periastron passage in units of the orbital period relative to MJD 58849 (2020), (7) parallax, and (8) system mass. The corresponding Bayesian priors placed on each parameter were (1) log uniform, (2) uniform, (3) sine, (4) uniform, (5) uniform, (6) uniform, (7) Gaussian ($9.6774\pm0.0429$\,mas), and (8) Gaussian ($2.71\pm0.14$\,$M_{\odot}$).

A random selection of orbits drawn from the posterior distributions is plotted in Figure~\ref{fig:orbit_tracks}, showing the plausible orbital trajectories relative to the resolved debris disk. The median and 1$\sigma$ credible intervals for the orbital elements derived from the $10^7$ OFTI samples are given in Table~\ref{tbl:orbit}. The marginalized distributions and their covariances for a subset of these parameters are shown in Figure~\ref{fig:corner_plot_OFTI}. We apply algebraic transformations to the OFTI samples to additionally derive marginalized distributions for the period using Kepler's third law, the periastron distance using the relationship $r_{\rm peri} = a(1-e)$, and the mutual inclination, $i_{\rm m}$, of the plane of the orbit of the companion relative to the debris disk using the formula:
\begin{equation}
    \cos i_{\rm m} = \cos i_{\rm disk} \cos i_{\rm b} + \sin i_{\rm disk} \sin i_{\rm b} \cos(\Omega_{\rm disk} - \Omega_{\rm b}).
\end{equation}
Because of the ambiguity of $\Omega_{\rm disk}$ we calculated $i_{\rm m}$ for each orbit for the two possible pairs of ($i_{\rm disk}, \Omega_{\rm disk}$) that are consistent with the southern part of the disk being behind the star ($i_{\rm disk} = 85$\,deg, $\Omega_{\rm disk} = 104$\,deg and $i_{\rm disk} = 95$\,deg, $\Omega_{\rm disk} = 284$\,deg; \citealp{Kalas:2015en}), selecting the smallest value to exclude retrograde orbits.

\subsection{Escape velocity}
\label{sec:esc_vel}
\begin{figure}
\includegraphics[width=\columnwidth]{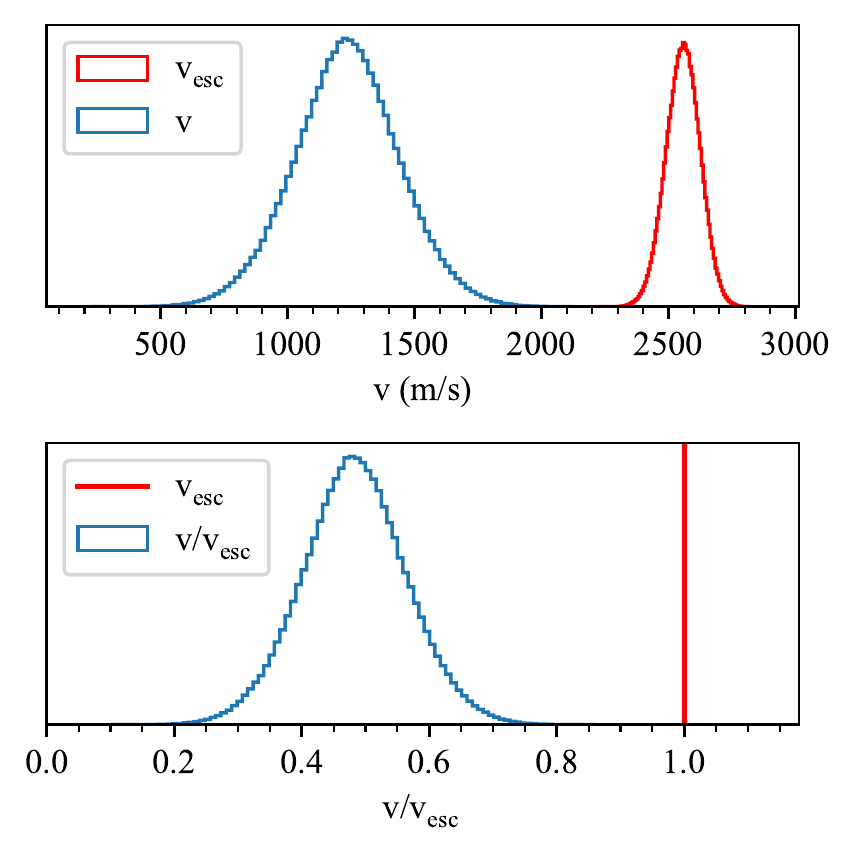}
\caption{Top: projected velocity of HD 106906 b relative to escape velocity of the system. Bottom: projected velocity of HD 106906 b divided by the escape velocity.\label{fig:esc_vel}}
\end{figure}

If HD 106906 b is bound to HD 106906, then the three-dimensional velocity of the companion must be lower than the gravitational escape velocity of the system. Radial velocity measurements of HD 106906 b relative to the host star have not yet been made, so we can only calculate a lower limit on the three-dimensional velocity. We estimated the projected linear velocity of the planet by calculating the instantaneous velocity of the companion at the midpoint of the orbit between the STIS and ACS epochs using the orbital element posterior distributions generated from OFTI. We found a projected velocity of the companion of $1.2\pm0.2$\,km\,s$^{-1}$, which is $48\% \pm 8$\% of escape velocity of the system (2.6\,km\,s$^{-1}$) estimated using a separation of 737 au and a mass of 2.71\,$M_\odot$ (Figure~\ref{fig:esc_vel}). While we cannot use this measurement to definitively conclude that the planet is gravitationally bound to HD 106906, a lower limit greater than the escape velocity would have been definitive evidence that the companion is unbound. Future radial velocity measurements of the planet will be necessary to constrain the three-dimensional relative velocity to demonstrate conclusively that HD 106906 b is gravitationally bound to HD 106906. 

\subsection{Differential parallax}
\label{sec:diff_plx}
The difference between the parallax of the star and planet should be negligible if the planet is gravitationally bound, less than 1 $\mu$as given a 700\,au orbit. We used the four astrometric measurements to fit a 10 parameter model for the relative astrometry of HD 106906 b following the procedure outlined by \citet{Nguyen:2020gv}. These 10 parameters included the standard five parameters ($\alpha$, $\delta$, $\pi$, $\mu_{\alpha^{\star}}$, and $\mu_{\delta}$) describing the astrometry of HD 106906 and five parameters ($\Delta\alpha$, $\Delta\delta$, $\Delta\pi$, $\Delta\mu_{\alpha^{\star}}$, and $\Delta\mu_{\delta}$) describing the relative position and motion of HD 106906 b. We used the same MCMC package as previously to sample the posterior distributions of the 10 parameters. We adopted Gaussian priors on the five parameters describing the astrometry of HD 106906 from the Gaia catalog values and uncertainties, and uniform priors for the others. We initialized 256 walkers near the maximum likelihood estimate and advanced them for 2000 steps, discarding the first 800 steps. We measured a differential parallax of $\Delta\pi = 3.66 \pm 5.73$\,mas, and a relative proper motion of $\Delta\mu_{\alpha} = 2.78 \pm 0.59$\,mas\,yr$^{-1}$, $\Delta\mu_{\delta} = 0.77 \pm 0.61$\,mas\,yr$^{-1}$. Our measurement was not particularly constraining; the differential parallax is consistent with zero but with a large uncertainty. Future astrometric monitoring of the system should greatly improve the uncertainties on this measurement.

\section{Discussion}
\label{sec:discussion}
\subsection{Is HD 106906 b perturbing the debris disk?}

\begin{figure}
\includegraphics[width=\columnwidth]{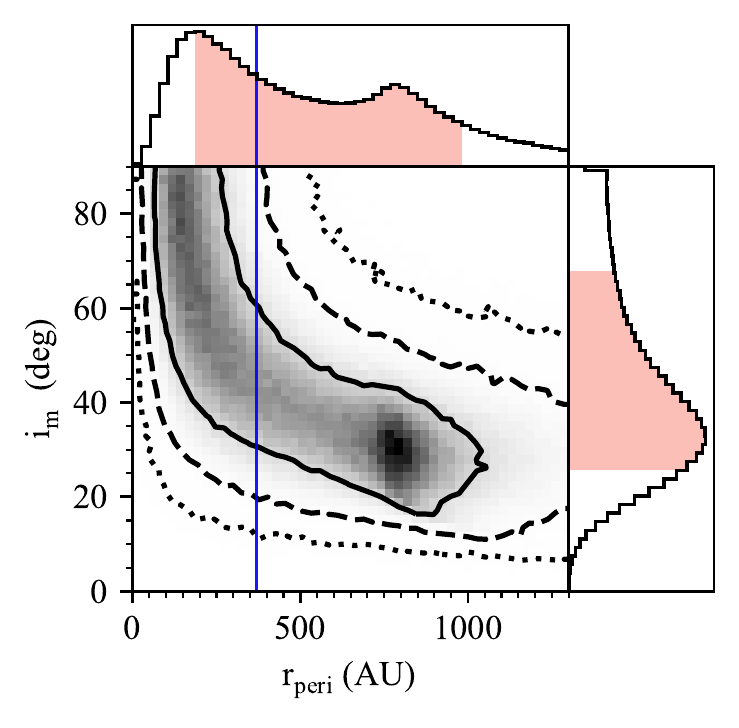}
\caption{The covariance between the mutual inclination of the inner debris disk and the orbital plane of the planet and the periastron distance of the planet. The corresponding marginalized distributions are also shown. The 1$\sigma$ credible interval for the marginalized distributions is highlighted in red. The 68th, 95th, and 99.5th percentile contours in the 2D covariance plot are plotted using solid, dashed, and dotted lines, respectively. The sensitivity-limited radius of the observed visible extent of the truncated eastern edge of the HD 106906 debris disk is denoted with a solid blue line. \label{fig:rperi}}
\end{figure}

Combining the orbital elements of HD 106906 b presented in Section~\ref{sec:orbits} with the orientation and inclination of the debris disk from \cite{Kalas:2015en} yielded a mutual inclination between the plane of the orbit of the planet and the inner system of either $36_{-14}^{+27}$\,deg or $44_{-14}^{+27}$\,deg (3$\sigma$ lower limit of 6 deg or 15 deg), depending on the true orientation of the orbital plane of the planet (Figure~\ref{fig:orbit_tracks}). We can confidently exclude a coplanar orbit (3-$\sigma$ lower limit is $13.8$\,deg or $5.1$\,deg depending on the value of $\Omega$) as well as a radial trajectory for the companion. The periastron distance and the mutual inclination between the orbital plane and the inner system are highly correlated (Figure~\ref{fig:rperi}). Eccentric orbits with periastron distances interior to the sensitivity-limited outer radius of the eastern side of the debris disk (370\,au) are highly misaligned with respect to the disk ($i_{\rm m} \gtrsim 40$\,deg), while more circular orbits with larger periastron distances are less misaligned ($i_{\rm m}\sim30$\,deg).

Previous studies have simulated the dynamical influence of an eccentric and misaligned planetary orbit on the debris disk \citep{Nesvold:2017ho}, assuming a small mutual inclination ($\sim$10\,deg) and a large eccentricity ($e\sim0.7$), corresponding to a small periastron distance ($\sim$200\,au). These simulations were able to qualitatively reproduce much of the observed morphology of the disk; however a thorough exploration of parameter space for the orbit of the companion is needed in future work. Our results suggest that an eccentricity as large as $e\sim0.7$ would require a very large mutual inclination of $\sim$60\,deg. This is larger than the maximum mutual inclination allowed based on the simulation and the lack of an observed vertical extent of the inner debris ring, potentially excluding some of the most eccentric solutions consistent with the astrometry presented in this work. The dynamical effect of a planet on an orbit with a lower eccentricity ($e\sim0.4$) but a larger mutual inclination ($i_{\rm m}\sim 30$\,deg), also consistent with our astrometric measurements, has not yet been explored.

\subsection{A Planet Nine analog?}
\begin{deluxetable}{lcccc}
\tabletypesize{\normalsize}
\tablecaption{Comparison of a selection of orbital elements for HD 106906 and other Solar System bodies.
\label{tbl:planet_nine}}
\tablehead{
\colhead{Param.} & \colhead{Unit} & \colhead{HD 106906} & \colhead{Planet 9} & \colhead{Detached KBOs} \\
& & Median ($\pm1\sigma$) & Range & Range}
\startdata
$P$            & yr & $15000_{-6400}^{+17000}$ & 8000--23000 & 1900--34000 \\
$a$            & au & $850_{-260}^{+560}$ & 400--800 & 160--1000 \\
$r_{\rm peri}$ & au & $510_{-320}^{+480}$ & 200--640 & 40--80\\
$e$            & \nodata & $0.44_{-0.31}^{+0.28}$ & 0.2--0.5 & 0.68--0.95\\
$i_{\rm m}$    & deg & $36_{-14}^{+27}$, $44_{-14}^{+27}$ & 15--25 & 4.2--33.5\\
\enddata
\end{deluxetable}
Table~\ref{tbl:planet_nine} compares our orbit constraints for HD 106906 b against the hypothetical Planet Nine \citep{Batygin:2019cq} and observations of 15 detached KBOs \citep{Trujillo:2020dk} demonstrating an approximate commensurability. With age $15\pm3$ Myr, HD 106906 b shows that a Planet Nine-like orbital architecture can be established early, though identifying the most likely pathway requires future work. One scenario is that the massive planet initially formed in a disk around the eccentric binary star, migrated into an unstable resonance with the stars, pumping the eccentricities and inclinations of the planet, and then the planet's periastron was raised beyond the central planetary region by passing stars \citep{Rodet:2017hr,DeRosa:2019ie}. If future observations can show that the binary stars, the disk, and HD 106906 b are all moving in a prograde direction, then this pathway would be favored as opposed to an alternate mechanism whereby HD 106906 b was captured from another star or as a free-floating planet. These same pathways are debated for the origin of a hypothetical Planet Nine \citep{Li:2016gb}. Also compelling is to investigate the theory of how the entire HD 106906 system will evolve in the future. Though there are differences between HD 106906 and the solar system, HD 106906 provides an empirical initial configuration for analytical and numerical experiments to determine how it may resemble the solar system after 4.6 Gyr of evolution.

\section{Conclusions}
We have presented the first measurement of the orbital motion of HD 106906 b about its host star. While the orbital period is long ($P\sim10^4$\,yr) a measurement of the trajectory of the planet was sufficient to exclude orbits that are coplanar with the inner debris disk \citep{Kalas:2015en,Lagrange:2016bh}---although this was assumed based on the current position of the planet relative to the inner system---and those that are moving outward in a radial trajectory that might be consistent with a planet in the process of being ejected from the system (e.g., \citealp{Rodet:2017hr}). The relatively large values for semi-major axis ($\sim$850 au), eccentricity ($\sim$0.44) and mutual inclination ($\sim$40 deg) are comparable to the estimated orbital properties of the hypothetical Planet Nine in our solar system.  This suggests that a Planet Nine-like orbital architecture can be established within the first $\sim$10 Myr of planet formation and subsequent dynamical evolution.  We measured a mutual inclination between HD 106906 b and the resolved debris disk of $36_{-14}^{+27}$\,deg or $44_{-14}^{+27}$\,deg, depending on the true orientation of the planet's orbit. HD 106906 is one of a small number of systems with an outer massive gas giant (e.g., $\pi$ Mensae; \citealp{Damasso:un,DeRosa:2020jc,Xuan:2020gf}) or substellar companion (e.g., GQ Lup; \citealp{2017ApJ...836..223W}) with a significant mutual inclination relative to the inner planet or inferred planetary system. We also measured a strong negative correlation between the periastron distance and the mutual inclination (Figure~\ref{fig:rperi}), suggesting that if the planet is indeed responsible for perturbing the disk, it likely has a significant mutual inclination with respect to the plane of the disk.

Previous work simulating the gravitational influence of the planet on the debris disk has shown that an eccentric and inclined planet external to the disk can reproduce the observed features of the disk \citep{Jilkova:2015jo,Nesvold:2017ho,Rodet:2017hr}. These studies were limited in the range of orbital parameter space they explored due to computational constraints, and they did not overlap with the parameters we estimate from our astrometric measurements. Repeating these simulations with mutual inclinations and eccentricities (or periastron distances) drawn from the locus shown in Figure~\ref{fig:rperi} may help further constrain the range of plausible orbits. It is likely that the most eccentric orbits consistent with the astrometry presented here would result in a significant disruption of the material at the smallest scales ($<100$\,au) in the inner system, inconsistent with near-infrared observations of the inner dust ring \citep{Kalas:2015en,Lagrange:2016bh}. A lower eccentricity, corresponding to a larger periastron distance and smaller mutual inclination, might be sufficient to minimize the impact on the inner dust ring while still perturbing the material seen at wider separations \citep{Kalas:2015en}.

Additional measurements of the relative astrometry between star and planet would be helpful to confirm the measurement of orbital motion presented here, and to refine the differential parallax measurement attempted in Section~\ref{sec:diff_plx}. If such measurements confirmed the results of this study, it is unlikely they would significantly reduce the uncertainties on the orbital parameters unless a significant amount of time had passed. At this point the most useful observation would be a measurement of the radial velocity of the planet; the systemic velocity of the primary has already been measured \citep{DeRosa:2019ie}, although to a relatively poor precision of $0.2$\,km\,s$^{-1}$ because it is an active, double-lined spectroscopic binary. A significant difference between the two would be conclusive evidence that the planet is not bound to the star (see Section~\ref{sec:esc_vel}), while a small but significant difference can measure the true orientation of the plane of the orbit, leading to a better constraint on the mutual inclination.

\acknowledgments
We thank Eric Nielsen, Gaspard Duch\^ene, Lea Hirsch, Thomas Esposito, and Chad Trujillo for discussions relating to this work, as well as the referee for their insightful comments. The authors were supported by NSF AST-1518332, NASA grants NNX15AC89G and NNX15AD95G, as well as HST-GO-14670 through a grant from STScI under NASA contract NAS5-26555. This work benefited from NASA’s Nexus for Exoplanet System Science (NExSS) research coordination network sponsored by NASA’s Science Mission Directorate. This research made use of the SIMBAD database and the VizieR catalog access tool, both operated at the CDS, Strasbourg, France. This work presents results from the European Space Agency (ESA) space mission Gaia. Gaia data are being processed by the Gaia Data Processing and Analysis Consortium (DPAC). Funding for the DPAC is provided by national institutions, in particular the institutions participating in the Gaia MultiLateral Agreement (MLA). Lastly, this research was based on observations made with the NASA/ESA Hubble Space Telescope, obtained from the data archive at the Space Telescope Science Institute. STScI is operated by the Association of Universities for Research in Astronomy, Inc., under NASA contract NAS 5-26555.
\facilities{HST (ACS, STIS, WFC3).}
\software{astropy \citep{2013A&A...558A..33A}, emcee, \citep{ForemanMackey:2013io}, matplotlib \citep{Hunter:2007ih}, orbitize \citep{Blunt:2019vq}, TinyTim \citep{Krist:2011kt}}
\clearpage

\bibliography{article}{}
\bibliographystyle{aasjournal}
\clearpage

\appendix
\section{Model comparison}
\begin{figure*}
\includegraphics[width=\textwidth]{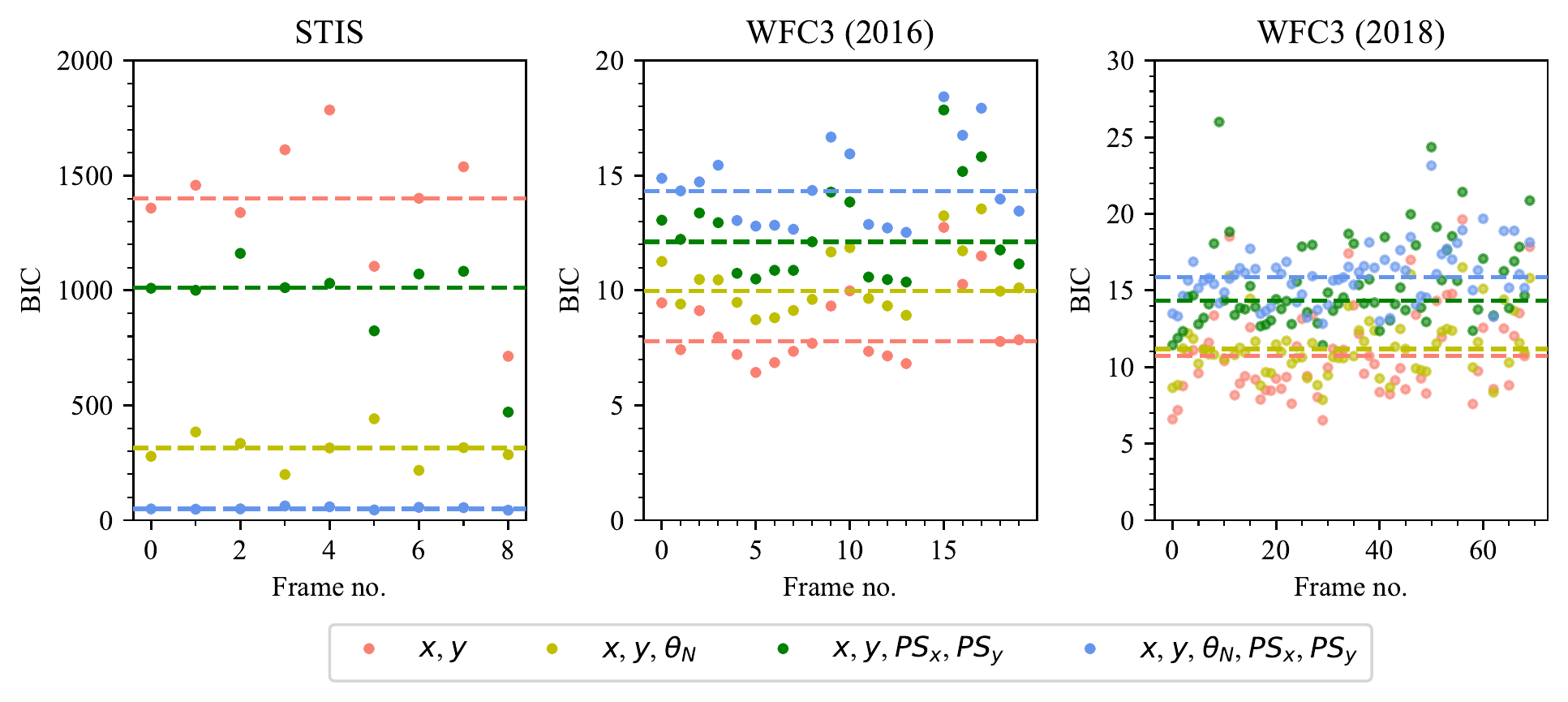}
\caption{BIC calculated for each frame of the STIS and WFC3 epochs using the four different permutations of free parameters we tested for our astrometric model.\label{fig:BIC}}
\end{figure*}

\restartappendixnumbering
We compared the four variants of our astrometric model by computing the BIC for each model applied to each image from the four epochs (Figure~\ref{fig:BIC}). A more complex model was only selected if the average $\Delta$BIC exceeded 10, as was the case for the STIS epoch.
\clearpage

\section{Verifying HST's astrometric calibration}
\label{sec:hst-calib}
\restartappendixnumbering
The significant offset between the derived position angle of north and the value derived from the FITS header for the STIS images prompted us to investigate the astrometric calibration of the three instrument configurations used in this study. We queried the archive for observations using a similar same instrument configuration. We cross-matched with the Gaia DR2 catalog to identify observations with three or more Gaia sources within the instrument field of view. In total, we obtained 1120 ASC/HRC, 1200 STIS/50CCD or STIS/50CORON, and 1113 WFC3/IR reduced datasets from the archive. We used the ``\_sx2'', ``\_x2d'', or ``\_drz'' files that are the end product of the observatory pipeline. We used the same algorithm we used for HD 106906 to measure the detector plate scale and orientation. We applied several automatic checks to reject sources that were either saturated or obscured by coronagraphic optical elements, as well as a manual screening of each observation to reject other problematic sources such as close binaries. We also rejected observations with an extremely high number of sources to avoid issues with source blending/confusion and potential Gaia systematic uncertainties in crowded regions.

\begin{figure*}
\includegraphics[width=\textwidth]{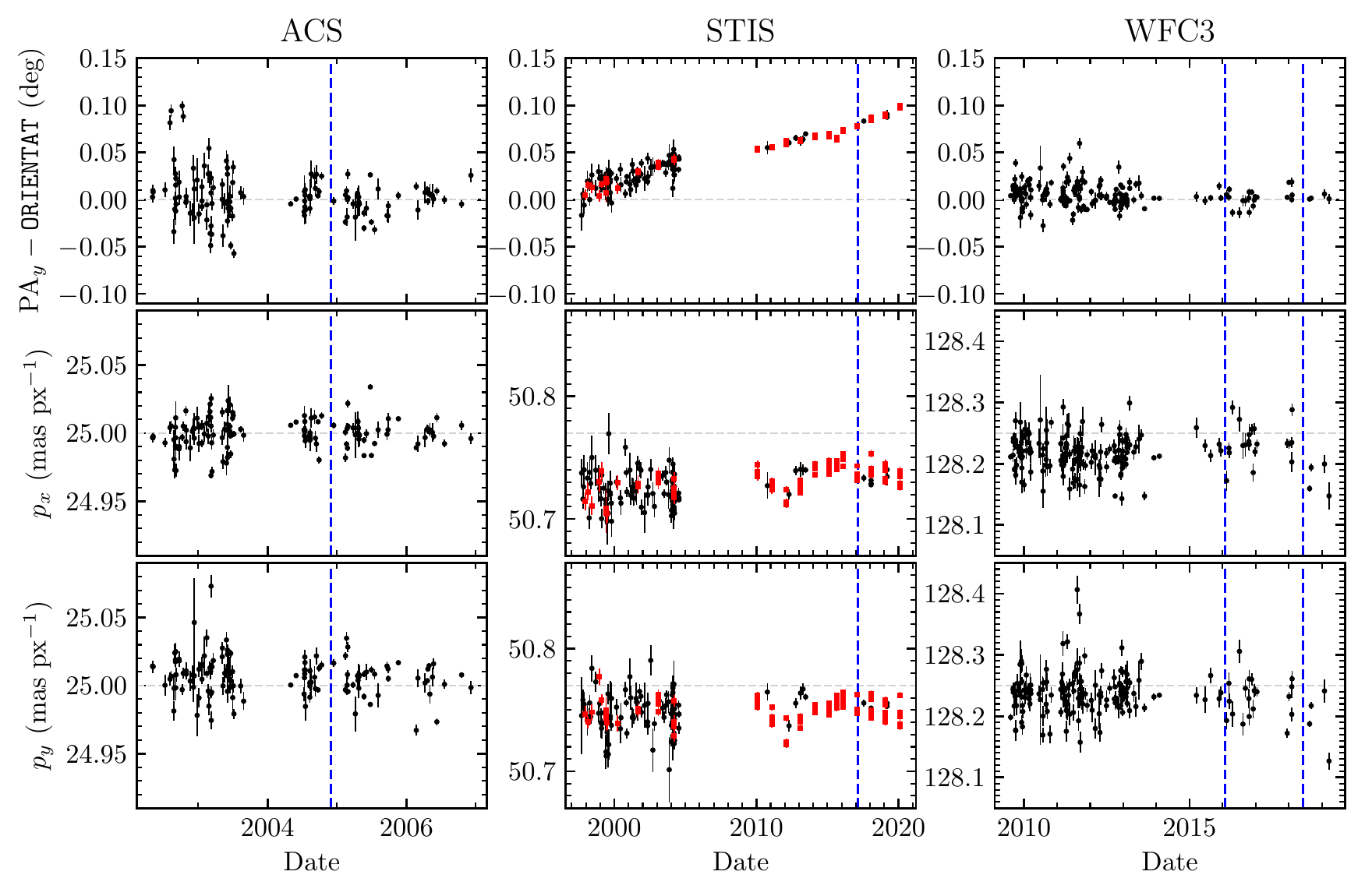}
\caption{Astrometric calibration of ACS HRC (left), STIS 50CCD/50CORON (middle), and WFC3 IR (right) measured using archival observations with at least 10 Gaia DR2 sources within the field of view as a function of date. Measurements derived from STIS observations of the globular cluster NGC 5139 are highlighted (red points). The top row shows the difference between the derived position angle of the detector $y$-axis and the \texttt{ORIENTAT} FITS header keyword, while the middle and bottom panels show the plate scales in the $x$ and $y$ directions. The nominal calibration values from the FITS header are denoted by the gray dashed lines. The date of HD 106906 observations are indicated by the vertical dashed blue line.\label{fig:validation}}
\end{figure*}

Our analysis of the ACS/HRC data yielded detector orientation and plate scales that were consistent to within a few hundredths of a degree for the orientation and 0.1\% for the plate scales with the calibration values reported in the FITS headers (Figure~\ref{fig:validation}, left column). The WFC3 data also had consistent orientation measurements (Figure~\ref{fig:validation}, right column), but with a 0.02\% smaller plate scale in both the $x$ and $y$ directions ($\sim$128.23 versus 128.25\,mas\,px$^{-1}$). This slight systematic offset of the plate scale is well within the uncertainties of our astrometric measurement of HD 106906 b. For the STIS data we measured a plate scale slightly lower (0.05\%) than the nominal value ($\sim$50.75 versus 50.77\,mas\,px$^{-1}$), and a significant offset between our measurement of the detector orientation and the FITS header values (Figure~\ref{fig:validation}, middle column). This offset is time-dependent, growing steadily at a rate of 0.004 deg\,yr$^{-1}$. This offset is also seen in observations of the globular cluster NGC 5139, a field that has been regularly observed with almost the same instrument configuration by the STIS instrument team to monitor detector performance. At the time of the HD 106906 observations in early 2017 this offset was approximately $0\fdg075$, consistent with the average difference between our measured orientation of the STIS images and the FITS header values ($0\fdg077\pm0\fdg002$). We note that this analysis was only performed with data obtained using the STIS 50CCD and 50CORON apertures; the other apertures were not tested and it cannot be assumed that the results presented here are valid for these other modes.
\clearpage
\clearpage

\section{Tables}
\label{sec:extra_tables}
\restartappendixnumbering

Here we tabulate the measured pixel position, tangent plane offset, and associated correlation coefficient for each Gaia catalogue background star for every image in all observing epochs (Table~\ref{table:dr2_catalogue}). We also present the best fit parameters for the primary star center  along with any relevant additional free parameters (e.g. the platescale, position angle of north) for each image in the ACS epoch in Table~\ref{tbl:acs-results}, the STIS epoch in Table~\ref{table:stis_5param}, the 2016 WFC3 epoch in Table~\ref{table:wfc3_2016_2param}, and the 2018 WFC3 epoch in Table~\ref{table:wfc3_2018_2param}.

\startlongtable
\begin{deluxetable*}{cccccccccc}
\tabletypesize{\footnotesize}
\label{table:dr2_catalogue}
\tablecaption{Pixel positions and tangent plane offsets for the sources used in this study.}
\tablewidth{0pt}
 \tablehead{
 \colhead{Instrument} & \colhead{Date} & \colhead{Filter} & \colhead{Image} & \colhead{{\it Gaia} DR2 Source} & \colhead{$x$} & \colhead{$y$} & \colhead{$\xi$} & \colhead{$\eta$} & \colhead{$\rho_{\xi,\eta}$}\\
 &&&&(607290-)&(px)&(px)&(mas)&(mas)&}
 \startdata
ACS & 2004-12-01 & f606w & j91711021 & 2994277277824 & $938.209\pm0.011$ & $319.208\pm0.011$ & $-8192.4\pm6.9$ & $8005.5\pm7.4$ & 0.350\\
 & & & & 2994268530304 & $1082.825\pm0.015$ & $731.929\pm0.013$ & $1433.7\pm15.1$ & $13213.5\pm26.9$ & 0.052\\
 & & & & 2994266748800 & $268.371\pm0.041$ & $72.535\pm0.042$ & $-11616.7\pm20.3$ & $-9485.3\pm13.0$ & 0.203\\
 & & & & (acs-3) & $1057.856\pm0.194$ & $713.013\pm0.196$ & \nodata & \nodata & \nodata\\
 & & & & (acs-4) & $808.014\pm0.162$ & $668.929\pm0.174$ & \nodata & \nodata & \nodata\\
 & & & & (acs-5) & $1078.365\pm0.060$ & $815.178\pm0.065$ & \nodata & \nodata & \nodata\\
 & & & & (acs-6) & $737.060\pm0.196$ & $1011.664\pm0.198$ & \nodata & \nodata & \nodata\\
 & & & & (acs-7) & $213.164\pm0.266$ & $999.423\pm0.265$ & \nodata & \nodata & \nodata\\
\enddata
\tablenotetext{a}{Image removed from analysis due to poor data quality.}
\end{deluxetable*}
\clearpage

\begin{deluxetable*}{rccc}
\tabletypesize{\normalsize}
\tablecaption{Fit parameters for the ACS epoch.\label{tbl:acs-results}}
\tablewidth{0pt}
\tablehead{\colhead{Property} & \colhead{Unit} & \colhead{No prior} & \colhead{With prior}}
\startdata
$x_0^{\rm acq}$ & px & $570.25_{-0.24}^{+0.23}$ & $569.91_{-0.20}^{+0.20}$\\
$y_0^{\rm acq}$ & px & $592.24_{-0.25}^{+0.25}$ & $592.20_{-0.20}^{+0.20}$\\
$\theta_{\rm N}$ & deg & $\equiv-80.95$ & $\equiv-80.95$\\
$p_x$ & mas px$^{-1}$ & $\equiv25.00$ & $\equiv25.00$\\
$p_y$ & mas px$^{-1}$ & $\equiv25.00$ & $\equiv25.00$\\
$\Delta x$ & px & $54.375_{-0.017}^{+0.017}$ & $54.375_{-0.017}^{+0.017}$\\
$\Delta y$ & px & $-69.194_{-0.034}^{+0.034}$ & $-69.194_{-0.034}^{+0.034}$\\
$\Delta \theta$ & deg & $0.0037_{-0.0020}^{+0.0020}$ & $0.0037_{-0.0020}^{+0.0020}$\\
$\chi^2_{\rm gaia}$ & \nodata & $2.84$ & \nodata \\
$\chi^2_{\rm align}$ & \nodata & $14.92$ & $14.92$\\
\hline
$x_0^{\rm coro}$ & px & $624.59_{-0.24}^{+0.23}$ & $624.25_{-0.20}^{+0.20}$\\
$y_0^{\rm coro}$ & px & $523.08_{-0.25}^{+0.25}$ & $523.04_{-0.20}^{+0.20}$\\
$\rho_{x_0y_0}$ & \nodata & $0.30$ & $0.00$\\
\hline
$\Delta x_{\rm 106b}$ & px & $205.34_{-0.25}^{+0.25}$ & $205.68_{-0.22}^{+0.22}$\\
$\Delta y_{\rm 106b}$ & px & $-197.73_{-0.26}^{+0.26}$ & $-197.70_{-0.21}^{+0.22}$\\
$\Delta \alpha^{\star}_{\rm 106b}$ & mas & $-5689.9_{-6.3}^{+6.3}$ & $-5690.3_{-5.3}^{+5.4}$\\
$\Delta \delta_{\rm 106b}$ & mas & $4291.2_{-6.5}^{+6.6}$ & $4299.7_{-5.4}^{+5.4}$\\
$\rho_{\rm 106b}$ & mas & $7126.6_{-5.5}^{+5.5}$ & $7132.1_{-5.4}^{+5.4}$\\
$\theta_{\rm 106b}$ & deg & $307.023_{-0.057}^{+0.058}$ & $307.076_{-0.043}^{+0.043}$\\
$\rho_{x_{\rm 106b}y_{\rm 106b}}$ & \nodata & $0.27$ & $-0.00$\\
$\rho_{\rho_{\rm 106b}\theta_{\rm 106b}}$ & \nodata & $-0.04$ & $0.01$\\
\enddata
\end{deluxetable*}

\begin{deluxetable*}{c|ccccc|c}
\tabletypesize{\normalsize}
\tablecaption{Fit parameters for the STIS epoch. \label{table:stis_5param}}
\tablewidth{0pt}
\tablehead{
\colhead{Image} & \colhead{$x_0$} & \colhead{$y_0$} & \colhead{$\theta_{\rm N}$} & \colhead{$p_x$} & \colhead{$p_y$} & \colhead{$\chi_{\nu}^2$} \\
&(px)&(px)&(deg)&(mas\,px$^{-1}$)&(mas\,px$^{-1}$)}
\startdata
od9t01010 & $565.74\pm0.02$ & $340.84\pm0.03$ & $84.769\pm0.002$ & $50.738\pm0.002$ & $50.766\pm0.003$ & 0.95 \\
od9t01020 & $555.90\pm0.02$ & $340.85\pm0.03$ & $84.770\pm0.002$ & $50.732\pm0.003$ & $50.757\pm0.003$ & 0.91 \\
od9t01030 & $230.61\pm0.03$ & $342.50\pm0.03$ & $84.764\pm0.002$ & $50.727\pm0.003$ & $50.757\pm0.003$ & 0.98 \\
od9t03010 & $566.19\pm0.02$ & $340.85\pm0.02$ & $69.866\pm0.003$ & $50.745\pm0.003$ & $50.767\pm0.003$ & 1.05 \\
od9t03020 & $556.30\pm0.02$ & $340.87\pm0.02$ & $69.865\pm0.002$ & $50.737\pm0.003$ & $50.759\pm0.003$ & 0.95 \\
od9t03030 & $231.09\pm0.03$ & $342.61\pm0.04$ & $69.863\pm0.003$ & $50.724\pm0.003$ & $50.755\pm0.004$ & 0.84 \\
od9t04010 & $566.15\pm0.02$ & $341.16\pm0.03$ & $54.865\pm0.003$ & $50.742\pm0.003$ & $50.763\pm0.003$ & 0.90 \\
od9t04020 & $556.28\pm0.02$ & $341.17\pm0.03$ & $54.864\pm0.002$ & $50.735\pm0.003$ & $50.757\pm0.003$ & 0.86 \\
od9t04030 & $231.04\pm0.04$ & $342.83\pm0.05$ & $54.866\pm0.003$ & $50.730\pm0.003$ & $50.751\pm0.005$ & 0.88 \\
\enddata
\end{deluxetable*}

\begin{deluxetable*}{c|cc|c}
\tabletypesize{\normalsize}
\tablecaption{Fit parameters for the WFC3 2016 epoch. 
\label{table:wfc3_2016_2param}}
\tablewidth{0pt}
\tablehead{
\colhead{Image} & \colhead{$x_0$} & \colhead{$y_0$} & \colhead{$\chi_{\nu}^2$}\\
& (px) & (px)}
\startdata
icytb0011 & $170.84\pm0.05$ & $173.12\pm0.04$ & 0.30 \\
icytb0041 & $170.82\pm0.05$ & $173.06\pm0.04$ & 0.15 \\
icytb0071 & $170.86\pm0.05$ & $173.12\pm0.04$ & 0.27 \\
icytb00a1 & $170.88\pm0.06$ & $173.08\pm0.04$ & 0.19 \\
icytb1011 & $188.82\pm0.06$ & $149.87\pm0.05$ & 0.26 \\
icytb1041 & $188.79\pm0.07$ & $149.89\pm0.05$ & 0.18 \\
icytb1071 & $188.81\pm0.06$ & $149.89\pm0.05$ & 0.23 \\
icytb10a1 & $188.79\pm0.06$ & $149.91\pm0.05$ & 0.27 \\
icytb0021 & $170.84\pm0.05$ & $172.98\pm0.04$ & 0.17 \\
icytb0051 & $170.85\pm0.05$ & $172.98\pm0.04$ & 0.29 \\
icytb0081 & $170.86\pm0.05$ & $172.97\pm0.04$ & 0.34 \\
icytb1021 & $188.82\pm0.06$ & $149.83\pm0.05$ & 0.27 \\
icytb1051 & $188.83\pm0.06$ & $149.83\pm0.05$ & 0.25 \\
icytb1081 & $188.83\pm0.06$ & $149.84\pm0.05$ & 0.22 \\
icytb0031 & $171.32\pm0.03$ & $173.78\pm0.03$ & 10.88\tablenotemark{a} \\
icytb0061 & $170.82\pm0.04$ & $173.10\pm0.03$ & 0.53 \\
icytb0091 & $170.86\pm0.04$ & $173.13\pm0.03$ & 0.36 \\
icytb1031 & $188.78\pm0.04$ & $150.02\pm0.04$ & 0.69 \\
icytb1061 & $188.79\pm0.04$ & $149.97\pm0.04$ & 0.32 \\
icytb1091 & $188.84\pm0.04$ & $149.99\pm0.04$ & 0.32 \\
\enddata
\tablenotetext{a}{Image removed from analysis due to poor data quality.}
\end{deluxetable*}
\clearpage

\startlongtable
\begin{deluxetable}{c|cc|c}
\tabletypesize{\small}
\tablecaption{Fit parameters for the WFC3 2018 epoch.
\label{table:wfc3_2018_2param}}
\tablewidth{0pt}
\tablehead{
\colhead{Image} & \colhead{$x_0$} & \colhead{$y_0$} & \colhead{$\chi_{\nu}^2$}\\
& (px) & (px)}
\startdata
icytb2011 & $154.74\pm0.06$ & $78.72\pm0.05$ & 0.14 \\
icytb2041 & $154.76\pm0.06$ & $78.68\pm0.05$ & 0.18 \\
icytb2071 & $154.73\pm0.06$ & $78.72\pm0.05$ & 0.32 \\
icytb20a1 & $154.64\pm0.06$ & $78.64\pm0.05$ & 0.50 \\
icytb3011 & $133.19\pm0.06$ & $75.93\pm0.04$ & 0.42 \\
icytb3041 & $133.17\pm0.06$ & $75.94\pm0.04$ & 0.31 \\
icytb3071 & $133.17\pm0.06$ & $75.92\pm0.05$ & 0.42 \\
icytb30a1 & $133.17\pm0.05$ & $75.84\pm0.04$ & 0.45 \\
\enddata
\tablenotetext{a}{Image removed from analysis due to poor data quality.}
\end{deluxetable}

\end{document}